\documentclass[twocolumn,aps,prx,floatfix,superscriptaddress,nofootinbib]{revtex4-2}
\usepackage{amsmath}
\usepackage{amsfonts}
\usepackage{amssymb}
\usepackage{graphicx}
\usepackage{color}
\usepackage{bm}
\usepackage{dsfont}
\usepackage{soul}
\usepackage{mathtools}
\usepackage{appendix}
\usepackage{bm}
\usepackage{mathrsfs}
\usepackage{esint}
\usepackage{slashed}
\usepackage{xcolor}
\usepackage{multirow}
\usepackage{wasysym}

\usepackage[final]{hyperref} 
\hypersetup{
	colorlinks=true,       
	linkcolor=blue,        
	citecolor=blue,        
	filecolor=magenta,     
	urlcolor=blue         
}

\usepackage{nccmath}
\usepackage{array}
\usepackage{hyperref}
\newcommand{\beq}{\begin{equation}}
\newcommand{\eneq}{\end{equation}}
\newcommand{\bea}{\begin{eqnarray}}
\newcommand{\eea}{\end{eqnarray}}

\begin{document}

\title{Post-quench relaxation dynamics of Gross-Neveu lattice fermions}

\author{Domenico Giuliano}
\affiliation{Institut f\"ur Theoretische Physik, Heinrich-Heine-Universit\"at, 40225 D\"usseldorf, Germany}
\affiliation{I.N.F.N., Gruppo collegato di Cosenza, 
Arcavacata di Rende I-87036, Cosenza, Italy}
\affiliation{Dipartimento di Fisica, Universit\`a della Calabria Arcavacata di 
Rende I-87036, Cosenza, Italy}
\author{Reinhold Egger}
\affiliation{Institut f\"ur Theoretische Physik, Heinrich-Heine-Universit\"at, 40225 D\"usseldorf, Germany}
\author{Bidyut Dey}
\affiliation{I.N.F.N., Gruppo collegato di Cosenza, 
Arcavacata di Rende I-87036, Cosenza, Italy}
\author{Andrea Nava}
\affiliation{Institut f\"ur Theoretische Physik, Heinrich-Heine-Universit\"at, 40225 D\"usseldorf, Germany}

\begin{abstract}
We study the quantum relaxation dynamics  for a lattice version of the one-dimensional (1D) $N$-flavor Gross-Neveu (GN) model after a Hamiltonian parameter quench. Allowing for a system-reservoir coupling $\gamma$, we numerically describe the system dynamics through a time-dependent self-consistent Lindblad master equation. For a closed ($\gamma=0$) finite-size system subjected to an interaction parameter quench, the order parameter dynamics exhibits oscillations and revivals. In the thermodynamic limit, our results imply that the order parameter reaches its post-quench stationary value in accordance with the eigenstate thermalization hypothesis (ETH). However, time-dependent finite-momentum correlation matrix elements equilibrate only if $\gamma>0$. Our findings  are consistent with the system being described by a pertinent Generalized Gibbs Ensemble (GGE) and, accordingly,  
highlight subtle yet important aspects of the post-quench relaxation dynamics of quantum many-body systems. 
\end{abstract}
\date{\today}
\maketitle

\section{Introduction}
\label{intro}
 
The continuous development and improvement of time-resolved spectroscopic 
techniques has triggered a remarkable increase of interest in the nonequilibrium 
dynamics of correlated electronic systems
 \cite{Giannetti2011,Graf2011,Smallwood2014,Peronaci2015,Caviglia2012,Nava2018}.   In general, 
there are several ways to study the nonequilibrium dynamics of quantum many-body systems. 
An effective and widely employed procedure is to prepare the system in a given 
equilibrium state (determined by an initial Hamiltonian) and subsequently, at time $t=0$, to perform a sudden quantum quench of one or several Hamiltonian system parameters. The post-quench dynamics then reveals characteristic time dependences of physically observable quantities. 
If the system allows for different equilibrium phases that are close in energy, the quench can induce  a time evolution across the phase boundary. In such cases, it is important to 
clarify the phase eventually reached at long times in a given quench protocol \cite{Lee2006,Andre2012,Sandri2015}. If both the quench parameters and the environment of the system can be controlled to high precision, the intriguing possibility opens up to engineer phases with 
desired properties at will \cite{Fu2014}.
Under suitable conditions, the quench dynamics can also lead to a dynamical phase transition (DPT) at some critical time where the system switches between different phases  \cite{Zvyagin2016,Heyl2018,Heyl2019,Dimeglio2020,Pellissetto2020,Rossini2020,Nava2023_s,Nava2024}. We note in passing that the presence of DPTs allows one to devise efficient protocols realizing so-called Pontus-Mpemba effects \cite{Nava2025b,Nava2025_s}.  

In this paper, we numerically study the post-quench relaxation dynamics for the paradigmatic example of  1D correlated lattice fermions with $N$ flavors and interaction parameter $g$, see also Ref.~\cite{Nava2025_s} for related work.
In the continuum limit 
this lattice model is equivalent to the $N$-flavor massless 1D GN model 
\cite{Gross1974,Affleck1982,Wolff1985,Pausch1991,Schnetz2004,Thies2006}, which develops an asymptotically free phase with dynamically generated fermion mass $m\ne 0$. 
Moreover, for $N=1$, our lattice model is equivalent to the model introduced in Refs.~\cite{Brazovskii1981,Brazovskii1984,Saxena1987} 
for describing the Peierls transition in 1D interacting polymers at half-filling. 
Specifically, we 
imagine that at times $t<0$, the system is prepared in an ordered phase with $m\ne 0$. At time $t=0$, we suddenly quench the interaction strength $g$ and let the system evolve with the post-quench Hamiltonian.   In order to account for the nonequilibrium dynamics of the order parameter, we employ a time-dependent self-consistent mean field (SCMF) approach \cite{Peronaci2015,Nava2023_s,Nava2024,Nava2025_s}. 

 In contrast to Ref.~\cite{Nava2025_s}, we here consider parameter quenches that do \emph{not} cause DPTs in the relaxation dynamics.  Instead, we are interested in clarifying the 
 combined effects of ETH and of a   system-environment coupling $\gamma\ge 0$, which may arise, e.g., from interactions between 
 quasiparticle excitations (typically neglected within the SCMF approximation), from particle exchange with a tunnel-coupled metallic substrate,
or from fluctuations of the order parameter in ordered phases \cite{Cui2019,Nava2024}, on the relaxation dynamics
 after a Hamiltonian parameter quench. In the $\gamma=0$ limit, the  system is expected to obey the ETH 
 \cite{Alessio2016}. As a consequence, in the thermodynamic limit $L\to \infty$, the order parameter characterizing a quench 
 should relax to a value corresponding to the average value of the corresponding observable computed in the 
 microcanonical ensemble, at an energy corresponding to the (conserved) total energy of the pre-quench state of the system. 
 Moreover, at finite $L$,  periodic revivals  (which should be suppressed at a finite $\gamma$) are expected to arise, 
 in the time dynamics of the order 
 parameter, in the closed system  \cite{Cardy2014,Rossini2020,Pellissetto2020}.  On 
 top of that, in determining  the possible equilibration of our system in the large-time limit, a key role is played by 
the possible integrability of our model Hamiltonian. The integrability of the 1D GN model has been  proven in the continuum version of 
the model \cite{Thies2014} and is somehow expected to persist in the lattice model Hamiltonian, as well. In integrable
systems, ETH is constrained by the presence of infinite conserved quantities, which makes the 
asymptotic state of the system described by a GGE, determined by 
requiring that all the conserved quantities take the value set in the initial state. For $\gamma = 0$, 
the asymptotic time evolution of our system is consistent with the onset  of the GGE. At variance, 
for $\gamma > 0$  our results also indicate that the time-dependent correlation matrix elements equilibrate,
which evidences how a true thermalization of the entire system to a stationary state is therefore possible only
for a finite coupling to the bath. 

In interacting systems such as ours, the time dependence must be pertinently handled 
when implementing approximations such as self-consistent mean field (SCMF) theory \cite{Peronaci2015,Mazza2017,Nava2023_s,Nava2024}. 
Moreover, a general treatment of the post-quench nonequilibrium dynamics requires addressing
dissipation effects due to a finite system-environment coupling $\gamma$. 
 As a result, the post-quench relaxation proceeds 
  from an initial nonequilibrium system state toward a final equilibrium state which is 
 determined by the post-quench parameters and by details of the
 system-environment coupling  \cite{Nava2021,Nava2023,Cinnirella2024,Cinnirella2025}. 
A time evolution toward a designated target state may then be achieved by engineering the quench parameters and/or the system-environment coupling, see, e.g., Refs.~\cite{Nava2024,Nava2025b}. 
Dissipation effects arising in the post-quench dynamics for $\gamma>0$ are described by
the Lindblad master equation (LME) \cite{Lindblad1976,Breuer2007,Fazio2025}.  Due to the time-dependent SCMF relation, this LME is effectively nonlinear and time-dependent.
However, since the problem studied below corresponds to quasi-free fermions, i.e., 
the Hamiltonian is quadratic (after imposing the SCMF approximation) and the Lindblad jump operators are linear in the fermion operators,
one can equivalently solve the LME in a simpler manner by switching to a closed set of differential equations for  the time-dependent correlation matrix
\cite{Fazio2025}. By numerically solving the latter equations, we obtain the dynamics of all physical quantities of interest, including the order parameter $m(t)$. We focus on the dependence of $m(t)$ on the quench amplitude, on the system size $L$, and on  $\gamma$.

In particular, we   first examine what happens in a closed   system with $\gamma=0$.
For relatively small quench amplitudes and at finite $L$, we find undamped oscillations 
in $m (t)$, witnessing the nonequilibrium character of the post-quench time evolution. At large 
enough quench amplitudes, so to have, in the initial state,  a nonnegligible fraction of 
excited quasiparticles at high enough energy to effectively vehiculate a damping of 
the order parameter,  an 
attenuation in $m(t)$ emerges, which seemingly mimics the onset of 
a dissipative dynamics.  Similar features have been reported for
the post-quench order parameter dynamics of $s$-wave superconductors \cite{Peronaci2015}.  However,  from the attenuation in $m(t)$, one cannot infer that the system globally evolves toward a stationary equilibrium state. 
Indeed, for finite $L$, we find recurrences in $m(t)$, where after a time interval $\Delta t \propto L$, $m(t)$ revives to a finite and large
value comparable to the initial one, see also Refs.~\cite{Cardy2014,Rossini2020,Pellissetto2020}. 
Even though extrapolation of our results to $L \to \infty$ would rule out such revivals in the thermodynamic limit, 
signatures for the lack of equilibration of the closed system  (corresponding to the onset 
of the GGE, for $t \to \infty$)  are hidden in the finite-momentum correlation matrix elements. We therefore conclude that for $\gamma=0$, the system does not relax to a true asymptotic equilibrium state.
This conclusion is not in contradiction to the ETH which either applies to a global 
order parameter like $m(t)$, or to the full dynamics of a local subsystem \cite{Alessio2016}.
For $\gamma>0$, however, all correlation matrix elements as well as $m(t)$ converge to their asymptotic steady-state values 
at $t\to \infty$, which in turn are determined  by the post-quench values of the system parameters. In particular, the periodic revivals in $m(t)$ observed for $\gamma=0$
are suppressed for $\gamma>0$.

The remainder of this paper is organized as follows.  In Sec.~\ref{modham}, we introduce the
GN lattice model and implement the SCMF approximation in order to study the equilibrium phase diagram of our model. 
In Sec.~\ref{LME_TDSC}, we discuss the self-consistent LME approach for the time-dependent quench problem and derive the corresponding dynamical equations for the correlation matrix elements. In Sec.~\ref{isopost}, we analyze the post-quench dynamics of the closed system ($\gamma=0$), whereas Sec.~\ref{pqc} studies the case $\gamma>0$.
In Sec.~\ref{concl}, we summarize our results and offer an outlook.
The Appendix provides details about our derivations as well as additional results. In App.~\ref{freen}, we derive the equilibrium free energy. In App.~\ref{LME_appe}, we derive the LME from a microscopic model describing quasiparticle tunneling between the system and a metallic substrate,  in App.~\ref{ananon} we derive the exact solution for the post-quench dynamics in the absence of self-consistency, and,  in App.~\ref{GGE}, we   investigate the large-time behavior of our system, 
in view of  the special role played  by quantum integrability  for $\gamma=0$ in combination with the ETH, see, for instance, Ref.~\cite{Yuzbashyan2016}.  
 
\section{Model and SCMF approximation}
\label{modham}

We consider a lattice Hamiltonian $H$ describing $N$ flavors of spinless fermions ($\alpha=1,\ldots,N$) on a $L$-site chain with periodic boundary conditions.
Using fermion annihilation operators $c_{j,\alpha}$ with  $j=1, \ldots, L$, fermions interact 
with a lattice displacement field $\{\Delta_j\}$ via the interaction strength $g>0$.  Assuming a vanishing chemical potential corresponding to the half-filled case and using the (real-valued) nearest-neighbor hopping amplitude $J$,  we study the Hamiltonian 
\begin{eqnarray}
H &=& - \sum_{j=1}^L (J + \Delta_j) \sum_{\alpha=1}^N 
\left[c_{j,\alpha}^\dagger c_{j+1,\alpha}^{} + c_{j+1,\alpha}^\dagger c_{j,\alpha}^{} \right]  \nonumber \\
&+& \frac{N}{2g} \sum_{j=1}^L \Delta_j^2 . 
\label{mh.1}
\end{eqnarray}
Below, we set the lattice constant to $a_0=1$.  Moreover, the energy unit will be set by $J=1$ throughout.
 Functionally integrating over the field $\{\Delta_j\}$ one recovers the (lattice version of the) 
``standard'' 1D GN Hamiltonian, $H_{\rm GN}$,  as \cite{Nava2025_s}
\begin{eqnarray}
H_{\rm GN}  &=& - J \sum_{j=1}^L   \sum_{\alpha=1}^N 
\left[c_{j,\alpha}^\dagger c_{j+1,\alpha}^{} + c_{j+1,\alpha}^\dagger c_{j,\alpha}^{} \right]  \nonumber \\
&+& \frac{g}{2N} \sum_{j=1}^L   \left\{ \sum_{\alpha=1}^N 
\left[c_{j,\alpha}^\dagger c_{j+1,\alpha}^{} + c_{j+1,\alpha}^\dagger c_{j,\alpha}^{} \right] \right\}^2    . 
\label{mh.1b}
\end{eqnarray} 
In the continuum limit, Eq.~\eqref{mh.1} corresponds to an 
$N$-flavor generalization of the model introduced in Refs.~\cite{Brazovskii1981,Brazovskii1984,Saxena1987} 
to describe the Peierls transition in 1D interacting polymers at half-filling. In this limit, it also corresponds to the 
1D $N$-flavor GN Hamiltonian \cite{Gross1974,Affleck1982,Wolff1985,Pausch1991,Schnetz2004,Thies2006,Nava2025_s}. 
Within the SCMF approximation, the displacement field is related to 
the fermion operators by a self-consistency relation,
\beq
\Delta_j  = \frac{g}{N}  \sum_{\alpha =1}^N  \left\langle c_{j,\alpha}^\dagger c^{}_{j+1,\alpha} + c_{j+1,\alpha}^\dagger 
c_{j,\alpha}^{} \right\rangle  , 
\label{mh.2}
\eneq 
where   $\langle \ldots \rangle$ denotes a quantum average over the fermionic many-body state $\rho$. 
We observe that only through Eq.~(\ref{mh.2}), fermion operators with different flavors are mixed with each other. Given a self-consistent solution for $\Delta_j$, the Hamiltonian (\ref{mh.1}) is therefore separable with respect to the flavor index. 

Let us first address the equilibrium phase diagram of this model.
According to the above discussion, one can write down energy eigenmode operators,
\beq
\Gamma_{\epsilon , \alpha} = \sum_{j=1}^L  u_{\epsilon, j}^* c^{}_{j,\alpha} , 
\label{mh.3}
\eneq
satisfying $[\Gamma_{\epsilon, \alpha} , H] = \epsilon \Gamma_{\epsilon, \alpha}$.  
The complex-valued wavefunction $u_{\epsilon , j}$ solves the time-independent 
lattice Schr\"odinger equation, 
\beq
 - (J+\Delta_j) \left[ u_{\epsilon , j+1} + u_{\epsilon , j-1} \right] = \epsilon u_{\epsilon , j },
\label{mh.4}
\eneq 
with $u_{\epsilon , j + L }= u_{\epsilon , j}$. Once Eq.~(\ref{mh.4}) has been solved, Eq.~(\ref{mh.2}) yields
\beq
\Delta_j = g  \sum_\epsilon \left( u_{\epsilon , j}^* u^{}_{\epsilon,j+1} + u_{\epsilon,j+1}^* u^{}_{j,\epsilon} \right) 
f (\epsilon), 
\label{mh.5}
\eneq
with the Fermi distribution function $f (\epsilon ) = 1/(e^{\beta \epsilon}+1)$ for $\beta = 1/k_B T$.  The temperature $T$ is eventually set by the environment when including a finite system-environment coupling $\gamma>0$, see Sec.~\ref{LME_TDSC}. In this section, we assume an infinitesimal but finite coupling $\gamma=0^+$. 

\begin{figure}
\centering    
\includegraphics[width=\linewidth]{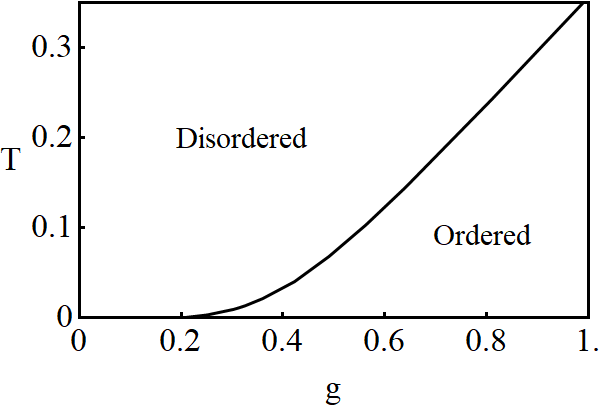}
\caption{Equilibrium phase diagram of the lattice GN model \eqref{mh.1} in the $g$-$T$ plane. 
Here $J=1$ sets the energy unit and we use $L=2000$ sites.
We assume the large-$N$ limit, where SCMF theory becomes exact. The solid curve marks the phase boundary between the ordered ($m\ne 0$) and the disordered ($m=0$) phase.  
As no significant changes are observed upon further increasing $L$, these results essentially 
correspond to the thermodynamic limit.}
\label{fig1}
\end{figure}

In the continuum version of SCMF theory for   the model in Eqs.~(\ref{mh.1},\ref{mh.1b}) taken at half-filling (zero chemical potential), it is well established that Eq.~(\ref{mh.2}) has spatially uniform solutions \cite{Brazovskii1981,Brazovskii1984,Saxena1987,Gross1974,Affleck1982,Wolff1985,Pausch1991,Schnetz2004,Thies2006}. In the lattice version, this corresponds to setting  
 \beq \label{deltadecomp}
      \Delta_j = \delta J + (-1)^j m,
 \eneq
 allowing both for a uniform ($\delta J$) and a staggered ($m$) component of the displacement field.  These two variables then serve as order parameters in the SCMF approach.
 At a finite chemical potential $\mu$, they can in principle be slowly varying functions of the site index $j$. 
However, numerically solving the SCMF equation directly in real space, without making any 
{\it a priori} assumption on the dependence on $j$, gives back the uniform solutions at half-filling \cite{Nava2025_s}.
For this reason, in the following we only consider spatially homogeneous solutions. 
 To account for $m\ne 0$, it is convenient to decompose the fermion operators into left- and right-movers. Switching to Fourier space, we write 
 \beq
 c_{j,\alpha}=  \frac{1}{\sqrt{L}} \sum_{0 \leq k \leq \pi}  e^{ i k j } c_{k,\alpha,1}
 + \frac{(-1)^j}{\sqrt{L}} \sum_{0 \leq k \leq \pi} e^{ikj} c_{k , \alpha ,2}, 
 \label{mh.6}
 \eneq
 where $c_{k,\alpha,1} = c_{k,\alpha}$ and $c_{k,\alpha,2} = c_{k+\pi , \alpha}$ with $0\leq k\leq \pi$ covering only half of the Brillouin zone. 
 Inserting Eq.~(\ref{mh.6}) into Eq.~(\ref{mh.1}) and defining 
 \beq\label{defcalJ}
 {\cal J} = J + \delta J,
 \eneq
 we arrive at
 \begin{eqnarray}
 H &=&    \frac{(m^2  + \delta J^2) L }{2 g} +  \sum_{0\leq k \leq \pi} \sum_{\alpha = 1}^N \left(c_{k,\alpha,1}^\dagger , c_{k,\alpha,2}^\dagger \right) \times \label{mh.7} \\
&\times&  
 \left( \begin{array}{cc} - 2 {\cal J}  \cos k  & - 2 i m \sin k   \\ 
 2 i m \sin k & 2 {\cal J} \cos k \end{array}\right)\left(\begin{array}{c} c_{k,\alpha,1}\\ 
 c_{k,\alpha,2} \end{array}\right)\nonumber .
 \end{eqnarray}
The fermionic quasiparticle spectrum is thus given by $\pm \epsilon_k$ with
\begin{equation}\label{dispferm}
\epsilon_k = 2 \sqrt{{\cal J}^2 \cos^2 k + m^2 \sin^2 k }.
 \end{equation}
In particular,  $m\ne 0$ opens a spectral gap $2|m|$ at $k=\pi/2$, where the SCMF 
 solution always yields $|m|<|{\cal J}|$.  
 The corresponding eigenmodes $\Gamma_{k,\alpha,\lambda=\pm}$ for energy $\pm \epsilon_k$ are given by  
 \beq
 \left( \begin{array}{c} \Gamma_{k,\alpha,+} \\ \Gamma_{k,\alpha,-} \end{array}\right) = \left(\begin{array}{cc} 
 \cos  \frac{\vartheta_k}{2} &  i \sin \frac{\vartheta_k}{2}   \\  
 i \sin  \frac{\vartheta_k}{2}  &   \cos \frac{\vartheta_k}{2} \end{array}\right)
 \left( \begin{array}{c} c_{k,\alpha,1}\\ c_{k,\alpha,2} \end{array} \right),
 \label{mh.8}
 \eneq
 where the angle $\vartheta_k$ is defined by
 \begin{equation}
 \cos \vartheta_k = -\frac{  2 {\cal J} \cos k}{\epsilon_k},\quad
 \sin \vartheta_k  = \frac{2 m \sin k}{\epsilon_k} . 
 \label{mh.9}
 \end{equation}
 From Eq.~(\ref{mh.5}), we obtain self-consistent expressions for $\delta J$ and $m$, 
 \begin{eqnarray}
 \delta J &=& \frac{4g}{L}\sum_{0 \leq k \leq \pi}  \frac{{\cal J}\cos^2 k}{\epsilon_k}  \tanh\frac{\beta \epsilon_k}{2} ,  \nonumber \\
 m &=&\frac{4g}{L} \sum_{0 \leq k \leq \pi} \frac{m \sin^2 k}{\epsilon_k}  \tanh  \frac{\beta \epsilon_k}{2} . 
 \label{mh.10}
 \end{eqnarray}
When complemented with the  free energy  in App.~\ref{freen},  Eq.~(\ref{mh.10})  determines the equilibrium phase diagram of our lattice model.  

Note that in a strictly 1D system ($N=1$), a finite coupling $g$ can never yield  an ordered ($m\ne 0$) phase at $T>0$ due to the Mermin-Wagner theorem. 
However, the SCMF approximation effectively implements a large-$N$ limit which becomes exact to lowest order in $1/N$.  In fact, by computing the free energy at finite $L$ and $N$, and afterwards sending $N\to\infty$ before taking the limit $L \to \infty$
 \cite{Gross1974,Affleck1982,Wolff1985,Pausch1991,Schnetz2004,Thies2006}, our system represents a 2D lattice model corresponding to an 
 array of $N$ coupled 1D chains of length $L$.  For this 2D case, finite-$T$ ordered phases are permitted.  In Fig.~\ref{fig1}, we show the 
phase diagram in the $g$-$T$ plane as obtained by numerical solution of the above equations for a system size $L$ large enough to lead to no additional changes on further increasing $L$. It exhibits the expected phase boundary between a low-$T$ asymptotically free phase with a dynamically generated mass gap ($m\ne 0$), and a high-$T$ trivial phase with $m=0$ \cite{Wolff1985}. The latter phase is qualitatively equivalent to  free relativistic fermions. 
 
\section{Time-dependent self-consistent Lindblad equation}
\label{LME_TDSC}

To describe  an  open  system coupled to its environment, we rely on the LME for the time-dependent density matrix $\rho (t)$ \cite{Breuer2007}. Specifically, we
employ Lindblad jump operators which are  proportional to  the quasiparticle creation and annihilation 
operators of the post-quench Hamiltonian \cite{Nava2021,Nava2023,Cinnirella2024,Nava2023_s,Nava2024,Nava2025_s}. 
Remarkably, the microscopic derivation of the LME for a chain tunnel-coupled to a metallic substrate yields exactly this form of the jump operators, see App.~\ref{LME_appe}.  In the LME for $\rho (t)$, we employ the time-dependent SCMF approximation which induces a time dependence of $m(t)$ and $\delta J(t)$, see Eq.~\eqref{deltadecomp},
and, consequently, of $H (t)$.  Since the resulting nonlinear and time-dependent LME for $\rho(t)$ describes quasi-free fermions, it is very convenient to solve it by
switching to the equivalent but simpler dynamical equations for the correlation matrix elements \cite{Fazio2025}. 
 These are the key quantities employed in computing time-dependent physical observables of the system. 

The time dependence of $H(t)$ also implies a time dependence of the single-particle eigenmodes, $\Gamma_{k,\alpha,\pm} (t)$, and of the corresponding eigenenergies
$\pm \epsilon_k (t)$.  Keeping time arguments implicit,
the LME takes the form, see App.~\ref{LME_appe}, 
\begin{eqnarray}\label{LME.1}
\frac{d \rho}{d t} &=& - i [ H , \rho  ] + \gamma  \sum_{\alpha = 1}^N\sum_{k,\lambda = \pm }  \biggl( f(-\lambda \epsilon_k  ) \times \\ \nonumber
&\times&
 \,{\cal D}[\Gamma_{k,\alpha,\lambda} ] \rho 
+ f (\lambda \epsilon_k)\,  {\cal D}[\Gamma_{k,\alpha,\lambda}^\dagger ]  \rho  \biggr),
\end{eqnarray}
with the Fermi function $f(\epsilon)$, the dissipator superoperator  
\beq
{\cal D}[\Gamma]\rho=\Gamma\rho\Gamma^\dagger - \frac12 \{\Gamma^\dagger \Gamma,\rho\},
\eneq
and the anticommutator $\{\cdot,\cdot\}$.  The jump operators used in Eq.~\eqref{LME.1} follow from the microscopic derivation in App.~\ref{LME_appe}.
Since the self-consistency condition is now enforced at every time step during the quench dynamics, we arrive at a nonlinear time-dependent LME.

We next define the correlation matrix elements ($a,a'=1,2)$
\begin{equation}\label{thetadef}
    \theta_{k,\alpha;(a,a')} (t) = {\rm Tr} \left[\rho (t) c_{k,\alpha,a}^\dagger c^{}_{k,\alpha,a'}\right].
\end{equation}\\
Since Eq.~(\ref{LME.1}) describes quasi-free fermions, one readily obtains equivalent linear first-order differential equations for the time dependence of these matrix elements, which are equivalent to Eq.~\eqref{LME.1} but much easier to solve numerically. We can thereby study relatively large systems.
Writing out the explicit time dependence, we find  
\begin{widetext}
\begin{eqnarray}\label{LME.3}
 \frac{d \theta_{k,\alpha;(1,1)} (t)}{d t} &=&
 2 m (t) \sin(k) 
\left[   \theta_{k,\alpha;(1,2)} (t) + \theta_{k,\alpha;(2,1)} (t) \right] 
- \gamma \theta_{k,\alpha ;(1,1)} (t) \\ \nonumber
&+&  \gamma f  ( \epsilon_k (t)  ) \cos^2 \left(\frac{\vartheta_k(t)}{2}  \right) +   \gamma 
f (-\epsilon_k (t)) \sin^2 \left(\frac{\vartheta_k (t)}{2} \right), \\ \nonumber
 \frac{d \theta_{k,\alpha ;(2,2)} (t)}{d t}&=& - \frac{d \theta_{k,\alpha;(1,1)} (t)}{d t} +
 \gamma \left[ 1- \theta_{k,\alpha;(1,1)}(t) - \theta_{k,\alpha ;(2,2) } (t) \right], \\
  \frac{d \theta_{k,\alpha ;(1,2)} (t)}{d t}&=&  -4i {\cal J} (t)   \cos (k) \,
\theta_{k,\alpha ;(1,2)} (t)  - 2 m(t)  \sin (k) \left[ \theta_{k,\alpha ;(1,1) } (t) - \theta_{k,\alpha ;(2,2)} (t) \right] \nonumber \\
&-& \gamma \theta_{k,\alpha ;(1,2) } (t) + \frac{i}{2} \gamma \left[ 1-2f   \left( \epsilon_k (t) \right)\right] \sin\vartheta_k (t ), 
\nonumber \\
  \frac{d \theta_{k,\alpha ;(2,1) } (t)}{d t}&=&  4i {\cal J} (t) \cos\left(k\right )   \,
\theta_{k,\alpha ;(2,1)} (t)  + 2 m  (t)  \sin (k) \left[  \theta_{k,\alpha ;(2,2) } (t) -
 \theta_{k,\alpha ;(1,1)} (t)  \right ] \nonumber \\ 
&-& \gamma \theta_{k,\alpha ;(2,1) } (t) - \frac{i}{2} \gamma \left[ 1-2f   \left( \epsilon_k (t) \right)\right] \sin\vartheta_k (t )
. \nonumber 
\end{eqnarray}
Here, $m (t)$ and ${\cal J} (t) = J + \delta J(t)$ have to be computed self-consistently at each time step, and
$\epsilon_k(t)$ and $\vartheta_k(t)$ follow from Eqs.~\eqref{dispferm} and \eqref{mh.9}, respectively, by substituting $m\to m(t)$ and ${\cal J}\to {\cal J}(t)$.
From Eq.~\eqref{mh.2}, we obtain the self-consistency relations
\begin{equation}
m (t) =- \frac{2ig}{NL}  \sum_{\alpha = 1}^N  \sum_{0\leq k \leq \pi}  \sin (k)  \left[\theta_{k,\alpha;(1,2)} (t) - 
\theta_{k,\alpha;(2,1)} (t) \right],\quad
\delta J (t)= \frac{2g}{NL}  \sum_{\alpha=1}^N   \sum_{0 \leq k \leq \pi}  \cos (k) 
  \left [\theta_{k,\alpha;(1,1)} (t) - \theta_{k,\alpha;(2,2)} (t) \right].
\label{LME.4}
\end{equation}
Moreover,  the time-dependent fermionic Hamiltonian can be written in the form 
\begin{equation}
H (t) =\sum_{\alpha = 1}^N\sum_{0 \leq k \leq \pi}   \epsilon_k (t)  \label{LME.5}  
 \left ( c_{k,\alpha,1}^\dagger , c_{k,\alpha,2}^\dagger \right) \left(\begin{array}{cc}
\cos \vartheta_k (t)  & - i \sin \vartheta_k (t)  \\
i \sin \vartheta_k (t)  & - \cos \vartheta_k (t)  \end{array} \right) \left(\begin{array}{c} c_{k,\alpha,1} \\
c_{k,\alpha,2}\end{array}\right)    .
\end{equation}
\end{widetext}

To initialize the system state at $t=0$, we evolve the open system for a large value of $\gamma$ in order  to minimize the preparation time, selecting the jump operators
such that the final state of this preliminary time evolution step corresponds to the desired  initial pre-quench state. Given the corresponding initial values 
$\theta_{k,\alpha;(a,a')} (0)$, we numerically evaluate the post-quench time evolution of the correlation matrix elements from Eqs.~\eqref{LME.3}. This also means that one has to simultaneously solve  the time-dependent self-consistency condition \eqref{LME.4} and to update the time-dependent Hamiltonian \eqref{LME.5}.

\section{Post-quench dynamics of closed system}
\label{isopost}

\begin{figure}[h]
    \centering    \includegraphics[width=\linewidth]{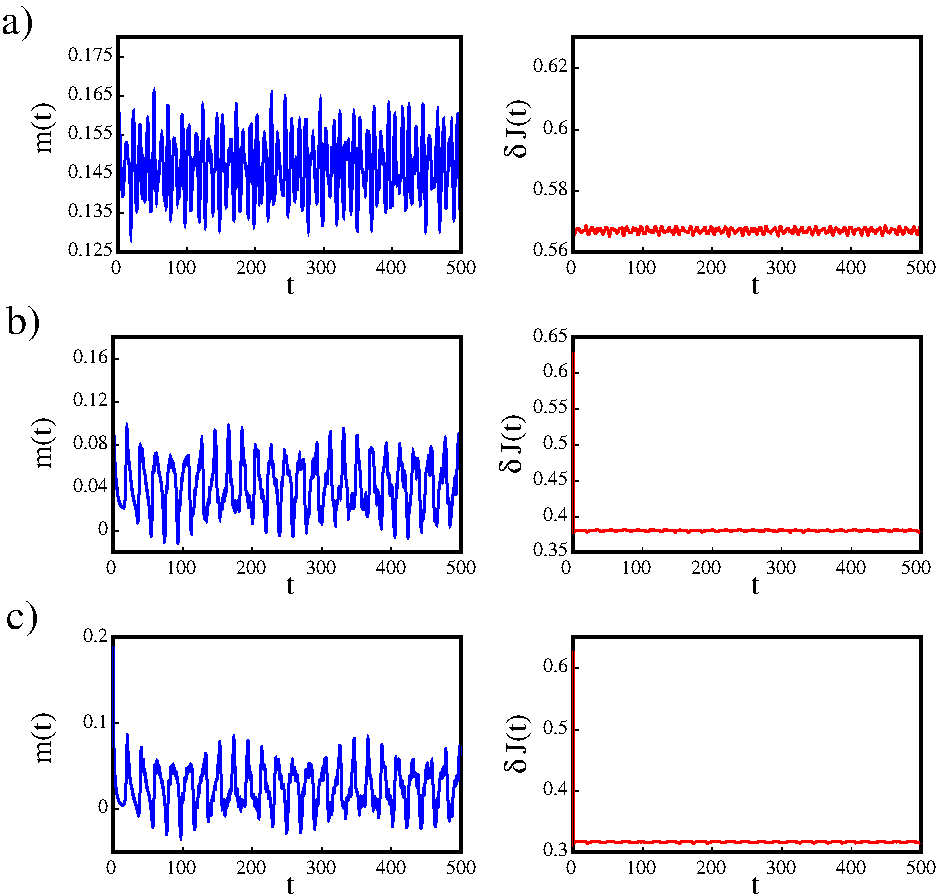}
    \caption{Post-quench order parameters $m(t)$ (left) and $\delta J(t)$ (right column)  
    vs time $t$ for a closed GN model in the large-$N$ limit where SCMF theory becomes exact. Units are determined by $J=1$.  The quench at $t=0$ is performed in the interaction strength, $g=g_i\to g=g_f$, with $g_i=1$ in all panels. Results were obtained by numerically solving Eqs.~\eqref{LME.3}, \eqref{LME.4} and \eqref{LME.5} for $L=100$, using (a) $g_f=0.9$, (b)  $g_f=0.6$, and (c) $g_f=0.5$. } 
    \label{fig2}
\end{figure}

We start by discussing the post-quench dynamics of closed systems (i.e., $\gamma = 0$).
 The SCMF-approximated model is described by a quasifree fermion Hamiltonian, 
with the only nonlinearity introduced by the self-consistent condition in Eq.(\ref{mh.2}). As a result, 
on giving up self-consistency, the model becomes quadratic and, therefore, integrable. Integrable
models are expected to partially satisfy the ETH, with the asymptotic state of the system 
corresponding to the GGE determined by the values of the (infinite) conserved quantities 
fixed by the initial state \cite{Alessio2016}. As it does not correspond to any conserved
quantity (see App.~\ref{GGE}),  the order parameter should   relax to the a stationary equilibrium value 
consistent with the GGE, in the thermodynamic limit. Remarkably, as we are going to 
highlight in the following, a similar dynamics is recovered when fully accounting for the 
time-dependent self-consistency in the system model Hamiltonian. 
On the numerical side, treating the time-dependent problem defined by Eqs.~\eqref{LME.3}, 
\eqref{LME.4}, and \eqref{LME.5} is more demanding than solving the  equilibrium problem in Sec.~\ref{modham}. For that reason, we show results for smaller system size $L$ below as compared to the phase diagram in Fig.~\ref{fig1}.  The pre-quench thermal initial state was determined by allowing for a large value of $\gamma$ with $k_B T=0.05 J$ for $t<0$.  
 Energy units are again set by $J=1$. 

We have performed time-dependent simulations for three different quench protocols. In all cases, we start from a state with pre-quench interaction strength $g_i=1$, corresponding to the ordered equilibrium phase region in Fig.~\ref{fig1}, and then perform a sudden parameter quench  $g_i\to g_f$ at time $t=0$, with 
$g_f<g_i$ still belonging to the ordered region. In Fig.~\ref{fig2}, we show $m(t)$ and $\delta J(t)$ for $t>0$, several $g_f$, and fixed size $L=100$. 

 \begin{figure}[h]
    \centering    
    \includegraphics[width=\linewidth]{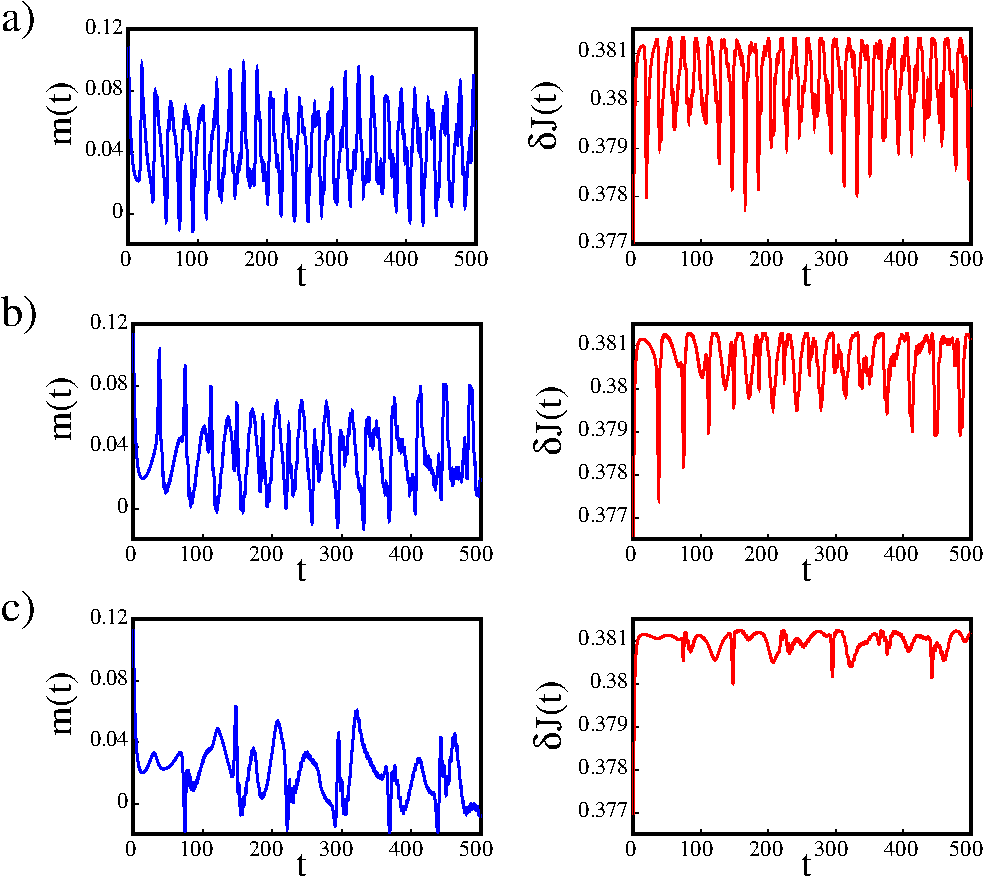}
    \caption{Post-quench order parameters $m(t)$ (left) and $\delta J(t)$ (right column) vs time $t$ for a closed GN system as in Fig.~\ref{fig2}(b) with $g_f=0.6$, but for different values of $L$. To better highlight the oscillations in $m(t)$ and $\delta J(t)$, we omit the initial drop at $t=0^+$ visible in Fig.~\ref{fig2} in this and the following figures. Results are shown for (a) $L=100$, (b) $L=200$, and (c) $L=400$.} 
    \label{fig3}
\end{figure}

 \begin{figure}[h]
    \centering    
    \includegraphics[width= \linewidth]{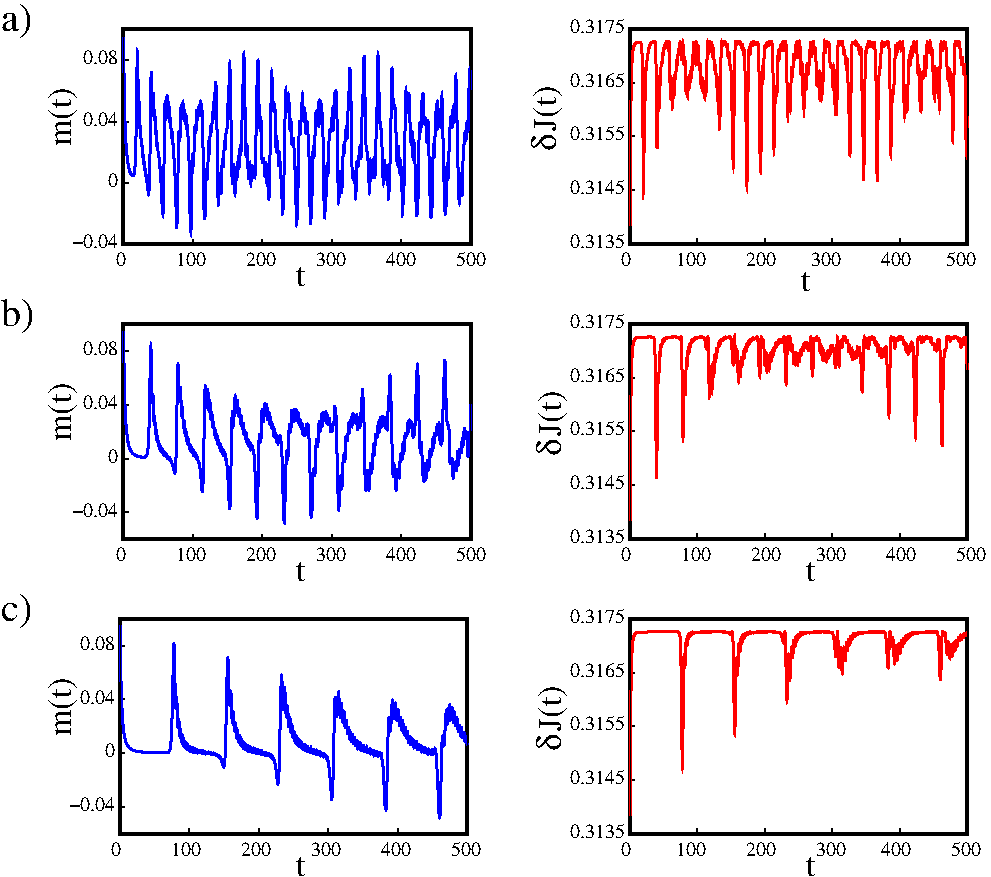}
    \caption{Post-quench dynamics of $m(t)$ (left) and $\delta J(t)$ (right column) for the closed GN model with different $L$ as in Fig.~\ref{fig3} but for $g_f=0.5$, see also Fig.~\ref{fig2}(c).  Results are shown for (a) $L=100$, (b) $L=200$, and (c) $L=400$.  } 
    \label{fig4}
\end{figure}

The quench at $t=0$ results in a nonequilibrium problem since the initial state is not a stationary state of the post-quench Hamiltonian. 
The resulting oscillatory behavior of both $m(t)$ and $\delta J (t)$  (where the oscillation amplitudes are much smaller)
is visible in all panels of Fig.~\ref{fig2}. Specifically, we find a sudden post-quench change in the average value of both order parameters at $t=0^+$, where the equilibrium post-quench value is approximately realized already.  Since $g_f<g_i$ in all panels, 
the rapid changes in $m(t)$ and $\delta J(t)$ imply initial drops in the magnitude of these order parameters. Subsequently, both 
order parameters oscillate around their corresponding equilibrium post-quench values, without sign of a damping mechanism reducing the oscillation amplitudes for $t\to \infty$. For finite size $L$, the oscillations are related to a finite revival time, $t_{\rm rev} \sim L / v$, where the velocity $v \sim 2{\cal J}$ characterizes elementary quasi-particle excitations  \cite{Cardy2014,Rossini2020,Pellissetto2020}. The scale $t_{\rm rev}$ is manifest in a slow periodic modulation of $m(t)$ and $\delta J(t)$, superimposed on fast oscillations. 

To further ground these observations, in Figs.~\ref{fig3} and \ref{fig4}, we demonstrate a direct relation between the oscillation frequencies and the system size $L$. Specifically, in Fig.~\ref{fig3}, we show $m(t)$ and $\delta J(t)$ for different $L$ with $g_f=0.6$, while in Fig.~\ref{fig4}, we analyze the corresponding case with $g_f=0.5$.
The time dependence of the order parameters is quite complex and determined by the superposition of several harmonics, where the relevant oscillation frequencies clearly depend on $L$. The undamped oscillatory time evolution in Figs.~\ref{fig3} and \ref{fig4} implies that we have persistent nonequilibrium states, without signature for a relaxation mechanism driving the system toward an asymptotic stationary state. 

Remarkably, for large quench amplitude $|g_f-g_i|$ and large $L$, see, e.g., Fig.~\ref{fig4}(c), 
the slow periodic modulations characterized by the time scale $t_{\rm rev}\sim L/v$ turn into sharp periodic revivals (\emph{aka} recurrences) 
in $m (t)$ and $\delta J(t)$.  For instance, $m(t)$ relaxes from the pre-quench value $m_i$ to the (here very small) ``final'' asymptotic value $m_f$ on a fast time scale, but then periodically revives to the pre-quench value $m_i$ at the times $t=nt_{\rm rev}$, with integer $n\ge 1$.  The revivals are qualitatively explained by the 
approximate expressions derived in App.~\ref{ananon}, where we give up self-consistency.  
In that case, an effective decoupling occurs between the parameter $2|m|$ (the single-particle mass gap) and the order parameters $m(t)$ and $\delta J(t)$ in Eq.~(\ref{LME.4}).
 
Specifically, we start from the pre-quench order parameters, $m_i$ and $\delta J_i$, determined from Eq.~(\ref{mh.10}). 
At $t=0$, we quench $g_i\to g_f$, with the corresponding stationary order parameters $m_f$ and $\delta J_f$ for $g=g_f$ obtained again from the equilibrium relation (\ref{mh.10}).  We also define $\epsilon_{k,i/f}$ as in Eq.~\eqref{dispferm} but 
with ${\cal J}\to {\cal J}_{i/f}=J+\delta J_{i/f}$ and $m\to m_{i/f}$. Without self-consistency, the Hamiltonian governing the post-quench dynamics is then time-independent. 
As detailed in App.~\ref{ananon}, it is convenient to define the time-dependent pseudovectors
\beq
\vec{\cal T}_{k,\alpha} (t) = \left(\begin{array}{c} \theta_{k,\alpha;(1,2)} (t)  + \theta_{k,\alpha;(2,1)} (t) \\ 
- i [ \theta_{k,\alpha;(1,2)} (t)  - \theta_{k,\alpha;(2,1)} (t) ] \\
 \theta_{k,\alpha;(1,1)} (t)  - \theta_{k,\alpha;(2,2)} (t)  
 \end{array}
 \right)  
 \label{pq0.1}
 \eneq
 subject to the initial condition  
\beq
\vec{\cal T}_{k,\alpha} (0) =- \left(\begin{array}{c}0 \\   \sin \vartheta_{k,i} \\  \cos  \vartheta_{k,i}   \end{array}\right) 
\label{pq0.4}
\eneq
with $\vartheta_{k,i/f}$ in Eq.~\eqref{mh.9} for $g\to g_{i/f}$.
For $t>0$, in the absence of self-consistency, one obtains decoupled equations of motion, 
 \beq
 \frac{d \vec{\cal T}_{k,\alpha}  (t)}{dt} = 2 \vec{\cal H}_k \times \vec{\cal T}_{k,\alpha} (t) , 
 \label{pq0.2}
 \quad
 \vec{\cal H}_k = \left(\begin{array}{c}0 \\ 2 m_f \sin k \\ - 2 {\cal J}_f \cos k  \end{array}\right). 
 \eneq
In Eq.~(\ref{anan.5}), we specify the analytical solution to Eq.~(\ref{pq0.2}) with the initial condition (\ref{pq0.4}). 
From Eq.~\eqref{LME.4}, this solution determines $m(t)$ and $\delta J (t)$ for $t>0$ as
\begin{eqnarray}
m (t) &=& \frac{2g_f}{NL}\sum_{\alpha = 1}^N \sum_{0 \leq k \leq \pi}\sin (k) {\cal T}_{k,\alpha}^y (t), \nonumber \\
\delta J (t) &=& \frac{2g_f}{NL} \sum_{\alpha = 1}^N  \sum_{0 \leq k \leq \pi} \cos (k) {\cal T}_{k,\alpha}^z (t)  . 
\label{pq0.6}
\end{eqnarray}
For $L$ sites (assuming even $L$), the quasi-momenta are $k_n = 2 \pi n/L$
with $n=1,\ldots , L/2$. Inserting Eq.~\eqref{anan.5} into Eq.~(\ref{pq0.6}), one finds     
\begin{equation}
m (t) = \bar{m} +  \delta m (t ),\quad 
\delta J(t) = \delta \bar{J}+ \delta \tilde J (t)  , 
\label{pq0.7}
\end{equation}
with
 \begin{eqnarray}
&& \bar{m} = \frac{16 g_f}{L}  \sum_{k_n} \frac{ [ {\cal J}_i {\cal J}_f \cos^2 k_n + m_i m_f \sin^2 k_n ] m_f
\sin^2 k_n }{\epsilon_{k_n,i} \epsilon_{k_n,f}^2},  \nonumber \\
&& \delta \bar{J}  = \frac{16 g_f}{L} \sum_{k_n} \frac{[ {\cal J}_i {\cal J}_f \cos^2 k_n + m_i m_f \sin^2 k_n ]{\cal J}_f
\cos^2 k_n }{\epsilon_{k_n,i} \epsilon_{k_n,f}^2},\nonumber \\ 
&&\delta m(t)  =     \frac{4 g_f}{L}  \sum_{k_n} \frac{  [m_f {\cal J}_i- m_i {\cal J}_f] \sin^2 (2k_n) 
  }{\epsilon_{k_n,i} \epsilon_{k_n,f}^2}   \cos [2 \epsilon_{k_n,f} t ], \nonumber \\
&& \delta \tilde J(t) = \frac{m_f}{{\cal J}_f}   \delta m (t) .  
\label{pq0.8}
\end{eqnarray}
  
\begin{figure}[t]
    \centering    
    \includegraphics[width= \linewidth]{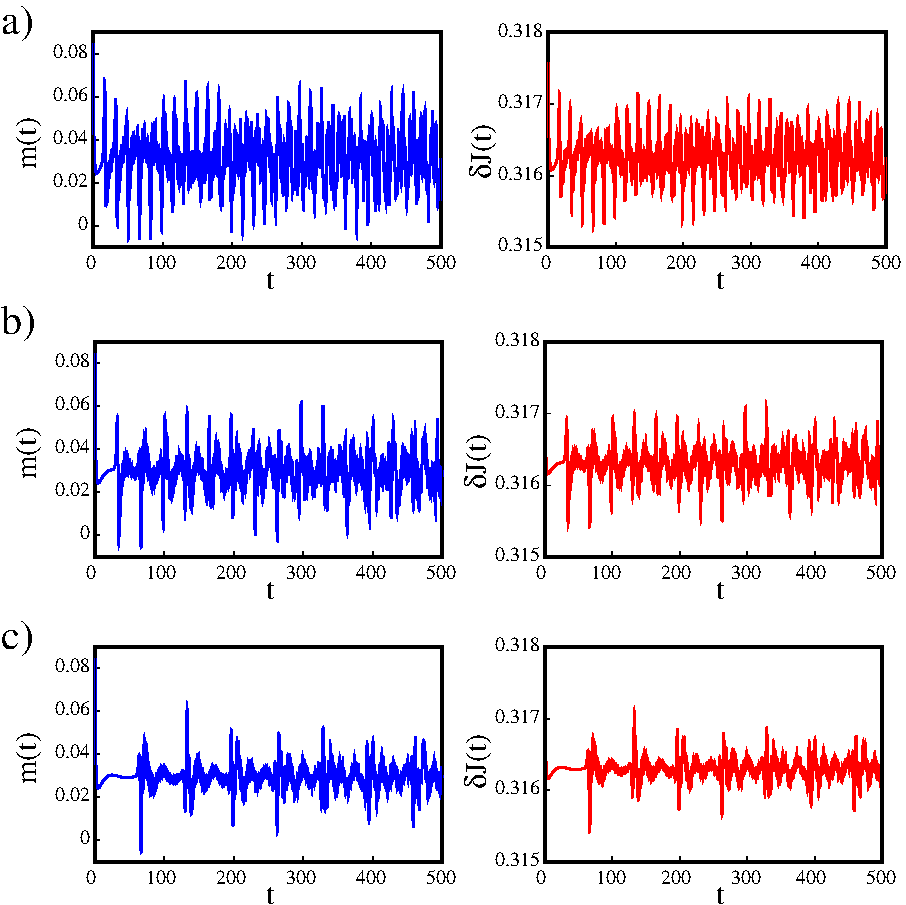}
    \caption{Post-quench dynamics of $m(t)$ (left) and $\delta J(t)$ (right column) vs time $t$ (in units with $J=1$) without enforcing time-dependent self-consistency.  As for the self-consistent counterpart in Fig.~\ref{fig3}, 
    we consider a quench from $g_i=1$ to $g_f=0.6$ for 
    (a) $L=100$, (b) $L=200$, and (c) $L=400$.    } 
    \label{fig5}
\end{figure}

 \begin{figure}
    \centering    
    \includegraphics[width= \linewidth]{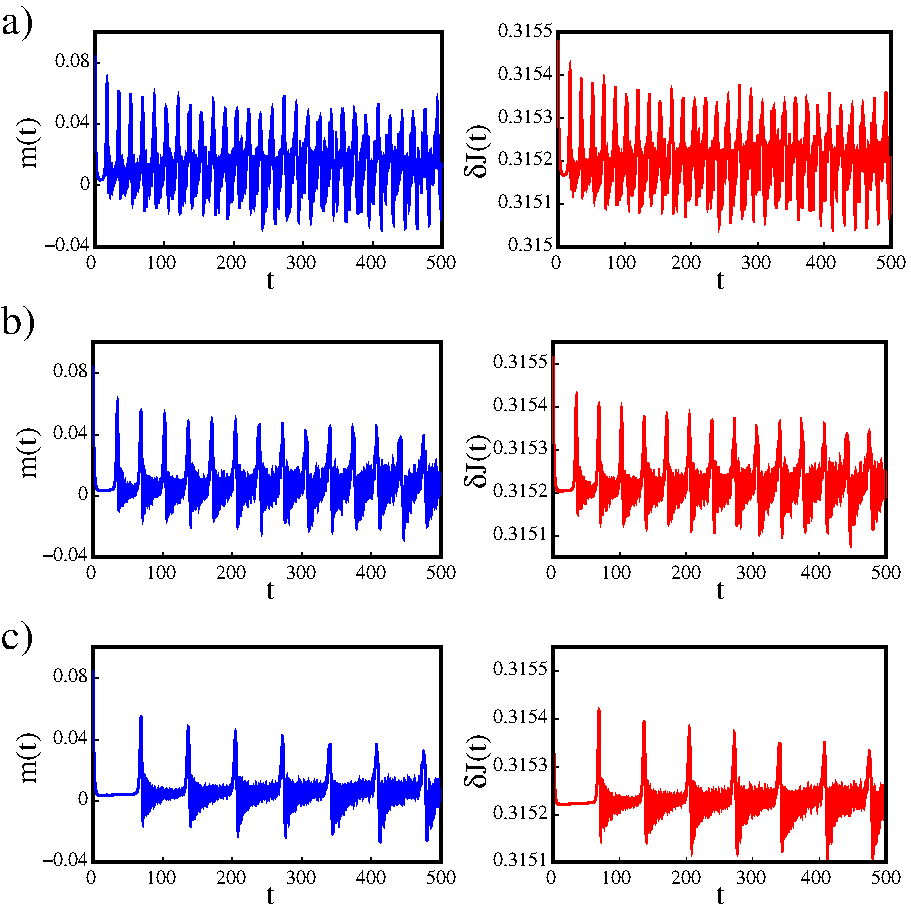}
    \caption{Post-quench dynamics of $m(t)$ (left) and $\delta J(t)$ (right column) vs time $t$ 
    without self-consistency.  As in Fig.~\ref{fig4}, we consider a quench from $g_i=1$ to $g_f=0.5$ for  (a) $L=100$, (b) $L=200$, and (c) $L=400$.  } 
    \label{fig6}
\end{figure}

In Figs.~\ref{fig5} and \ref{fig6}, we show the non-self-consistent counterparts to Figs.~\ref{fig3} and \ref{fig4}, respectively,
where $m(t)$ and $\delta J(t)$ are obtained from Eqs.~\eqref{pq0.7} and \eqref{pq0.8}.
Both the self-consistent and the non-self-consistent version show a similar scaling of the oscillation frequencies with $L$, with comparable results for $m(t)$ and $\delta J(t)$ in both approaches.   While this observation supports our subsequent use of the non-self-consistent approach for estimating 
 $\delta m(t)$  and $\delta \tilde J (t)$, see Eq.~\eqref{pq0.7}, in the asymptotic long-time limit, it is worth noting that the time-dependent SCMF method 
  determines the asymptotic values $m_f$ and $\delta J_f$ by itself. In the non-self-consistent variant, those values must be computed separately and then 
 inserted by hand into Eq.~\eqref{pq0.2}.  Nonetheless, Eq.~(\ref{pq0.8}) provides a reasonably good description for the oscillations 
 in $m(t)$ and $\delta J(t)$. In particular, the scaling of the relevant harmonic modes with $L$ can be extracted from this analytical approach. For large quench amplitude and large $L$, see, e.g.,  Figs.~\ref{fig4}(c) and \ref{fig6}(c), $m_f$ is very small and $m(t)$ exhibits sharp periodic revivals after a monotonic relaxation from $m_i$ to $m_f$.  The revivals  periodically (and approximately) recover the post-quench value again, $m(nt_{\rm rev})\approx m_i$. 
 
For $L \to \infty$, we instead have $t_{\rm rev}\to \infty$, and there are no revivals anymore.  The systems then seems to 
undergo a true relaxation dynamics towards a stationary equilibrium state. This is also seen from the non-self-consistent solution (\ref{pq0.8}) for, say, $m(t)$. 
As we discuss in detail in Appendix \ref{GGE}, the true nature of the 
asymptotic state can be inferred by putting together the ETH with the integrability of the non-self consistent 
Hamiltonian (which is simply quadratic and, therefore, integrable). Specifically, one readily shows that the asymptotic state is described by the density matrix $\rho_{\rm GGE}$, given by 
\beq
\rho_{\rm GGE}= \frac{e^{-\sum_{k,\alpha} \lambda_{k,\alpha} {\cal C}_{k,\alpha}}}
{{\rm Tr} [ e^{-\sum_{k,\alpha} \lambda_{k,\alpha} {\cal C}_{k,\alpha}} ]}  , 
\label{apn.5x}
\eneq
with 
\begin{eqnarray}
{\cal C}_{k,\alpha} &=& -i \sin (\vartheta_{k,f} ) \{c_{k,\alpha,1}^\dagger c_{k,\alpha,2} - c_{k,\alpha,2}^\dagger c_{k,\alpha,1} \}
\nonumber \\
&+& \cos (\vartheta_{k,f} ) \{c_{k,\alpha,1}^\dagger c_{k,\alpha,1} - c_{k,\alpha,2}^\dagger c_{k,\alpha,2} \} ,
\label{apn.6x}
\end{eqnarray}
and the $\{\lambda_{k,\alpha}\}$ being Lagrange multipliers that make the average value of ${\cal C}_{k,\alpha}$ equal to the one in the initial state. Equations (\ref{apn.5x},\ref{apn.6x}) define a nonequilibrium state, see
App.~\ref{GGE} for details. The striking qualitative similarity between  the results obtained with the self-consistent and with the non-self-consistent approach eventually lead us to conclude that a similar GGE state is determined by the asymptotic time evolution of the model once the full time-dependent SCMF procedure has been implemented.

To recover $\delta m(t)$ in the thermodynamic limit, we 
start from the finite-$L$ expression for $\delta m(t)$ in Eqs.~(\ref{pq0.8}) and 
substitute $\frac{1}{L} \sum_{k_n}$
(with $k_n = \frac{2 \pi n}{L}$ and $n=0,\ldots ,\frac{L}{2}-1$) with the integral $\int_0^\pi \: \frac{dk}{2\pi}$. 
Consistent with the fact that we are mainly interested in the asymptotic ($t \to \infty$) 
behavior, we expand the argument of the integral around the energy minimum by setting $k=\frac{\pi}{2}+q$ and retaining only the leading nontrivial contributions in $q$. This requires introducing a high-momentum 
cutoff $\Lambda$ with $0 \leq |q| \leq \Lambda$, eventually sending $\Lambda \to \infty$.      
Setting $m_f \approx 0$, and retaining, as discussed above,  only long-wavelength fermion excitations close to the band minimum, $\delta m(t)$ can be simplified, for $t\to \infty$, to 
\begin{eqnarray}
\delta m (t ) &\approx& - g_f [m_i {\cal J}_f - m_f {\cal J}_i]  \int_{-\infty}^\infty \frac{d q}{4 \pi {\cal J}_f^2}   
 \frac{e^{4 i q {\cal J}_f t}}{\sqrt{{\cal J}_i^2 q^2 + m_i^2 } }   \nonumber \\
&=& - \frac{g_f[m_i {\cal J}_f - m_f {\cal J}_i] }{2 \pi {\cal J}_i {\cal J}_f^2}  K_0 \left( \frac{4 {\cal J}_f m_i t}{{\cal J}_i}\right) \nonumber \\
&\sim& \sqrt{\frac{ \pi {\cal J}_i}{8  {\cal J}_f m_i t}} 
\exp \left(- \frac{4 {\cal J}_f m_i t}{{\cal J}_i}\right), 
\label{pq0.10}
\end{eqnarray} 
with the modified Bessel function $K_n(u)$ of the second kind. The last step in Eq.~\eqref{pq0.10} holds for $t\gg {\cal J}_i/[4 {\cal J}_f m_i]$, highlighting the 
exponential decay of the post-quench oscillations in $m(t)$ and $\delta J(t)$. 
 The order parameter is just a specific combination of real-space correlation functions, $m_j=\frac{1}{NL}\sum_{\alpha=1}^N \sum_{0 \leq k \leq \pi}e^{-ik(j-j')} 
\{\theta_{k;\alpha;(1,1)} (t) + (-1)^{j-j'} \theta_{k;\alpha;(2,2)} (t) +
(-1)^{j} \theta_{k;\alpha;(1,2)} (t)+(-1)^{j'} \theta_{k;\alpha;(2,1)} (t) \}$ 
for $j=j'$, with the correlation matrix elements $\theta_{k;\alpha;(a,a')}(t)$
defined in Eq.~(\ref{thetadef}). Using arguments similar to the ones 
leading to Eq.~(\ref{pq0.10}), we infer an exponential decay of  
correlations in time and in real space, with a typical length (or time)  $\propto {\cal J}_i/4m_f$. 
Using  Eq.~(\ref{anan.6}), a similar decay of correlations in real space is 
expected for the asymptotic stationary state reached for the open system at
$t\to \infty$. Just as the order parameter, the result for the real-space 
correlations is consistent with the ETH, which  applies to a 
global order parameter like $m(t)$, or to the full dynamics 
of  local subsystems \cite{Alessio2016}.  

\begin{figure}
    \centering    
    \includegraphics[width=0.8\linewidth]{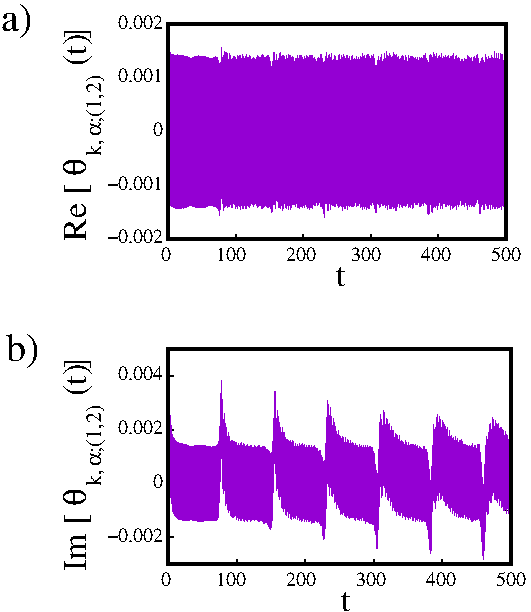}
    \caption{Post-quench dynamics of the (a) real part  and (b) imaginary part  
     of the finite-momentum correlation matrix element $\theta_{k_*,\alpha;(1,2)}$ vs time $t$ for the closed GN model, with $J=1$ and arbitrary $\alpha$. As in Fig.~\ref{fig3}(c), we use $g_f=0.5$ and $L=400$, and define the momentum scale 
    $k_* = 15 \frac{2 \pi}{L}$. These results were obtained from Eqs.~\eqref{LME.3} and \eqref{LME.4}.  } 
    \label{fig7}
\end{figure}
   
\begin{figure}
    \centering  
    \includegraphics[width=0.7\linewidth]{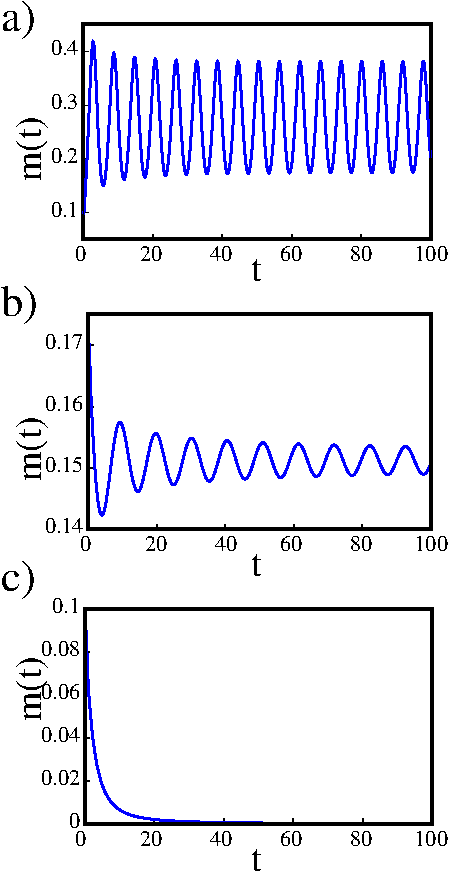}
    \caption{Post-quench dynamics of $m(t)$ for the closed GN model with $J=1$ for
    (a) $g_i=0.6, g_f=1.5$, (b)  $g_i=1, g_f=0.9$, and (c) $g_i=1, g_f=0.5$. Consistent with the numerical values of $m_i/\hat{m}_f$ (see main text), the dynamics of $m(t)$ corresponds to  regimes A, B, and C of Ref.~\cite{Barankov2006}, respectively. } 
    \label{new_figure}
\end{figure}

Similar apparent relaxations of order parameters after a large parameter quench have been reported for other closed many-body systems before, see, e.g., Refs.~\cite{Peronaci2015,Nava2023_s,Das2020}. 
In particular, our way of dealing with the post-quench 
dynamics without any coupling to a bath parallels the analysis performed in Refs.~\cite{Barankov2004,Barankov2006}, to which our framework is then formally equivalent.  In those papers, three different regimes were identified, depending on the ratio between the pre-quench value of the order parameter, $m_i$, and the value of the order parameter in the equilibrium state determined by the post-quench parameters, $\hat{m}_f$. 
 Specifically, for $m_i/\hat{m}_f< e^{-\pi/2}$ (regime A), 
 there is no apparent asymptotic attenuation of the order parameter.
 On the other hand, for $e^{-\pi/2} \leq m_i/\hat{m}_f 
 \leq e^{\pi/2}$ (regime B), the $t>0$ oscillations between
 the two limiting values $m_{\rm min}$ and $m_{\rm max}$ become damped. Due to the  Landau damping mechanism, for $t\to \infty$, $m (t)$ flows to an asymptotic value $\hat m_f$, which does not monotonically depend on $m_i/\hat{m}_f$. Finally, for 
 $e^{\pi/2} < m_i/\hat{m}_f$ (regime C), 
 the damping gets stronger, eventually washing out the post-quench oscillations in $m (t)$ and leading to $\hat m_f=0$. In Fig.~\ref{new_figure}, we show the post-quench time 
 evolution of $m(t)$ for those three cases. In particular, regime A corresponds to Fig.~\ref{new_figure}(a),  regime B to Fig.~\ref{new_figure}(b), 
 and regime C to Fig.~\ref{new_figure}(c). The curves for $m(t)$ demonstrate the consistency between our findings and those of Ref.~\cite{Barankov2006}.
   
In any case, we emphasize that the intrinsic post-quench dynamics features a decoupling between the global order parameter relaxation and the flow of arbitrary correlation matrix elements toward a stationary equilibrium state. In fact, for a true relaxation
to an equilibrium state, we expect a relaxation of all matrix elements  $\theta_{k,\alpha;(a,a')}(t)$ to a stationary value in the limit $t\to \infty$.  However, this is not observed for the presently studied closed systems as we illustrate 
in Fig.~\ref{fig7} for a finite-momentum correlation matrix element subject to the same protocol as in Fig.~\ref{fig3}(c).  Here, we again take self-consistency into account.
Such matrix elements directly contribute to the order parameters, see Eq.~(\ref{LME.4}). 
We observe  unattenuated oscillations of the real and imaginary parts of this matrix element persisting for arbitrarily long times. A comprehensive understanding of the full nonequilibrium time evolution of the closed system 
therefore cannot rely on a few global observables like $m(t)$ and/or $\delta J(t)$. The relaxation of these order parameters is 
caused by a superposition of many harmonics in the thermodynamic limit $L \to \infty$ which effectively undergo dephasing at long times, in accordance with the ETH \cite{Alessio2016}. The underlying nonequilibrium nature of the state is then encoded in the persistent unattenuated oscillations of the finite-momentum harmonics of order parameters, see Fig.~\ref{fig7}.
 As discussed in Sec.~\ref{pqc}, only for $\gamma>0$, a true relaxation to a stationary equilibrium state occurs, where the oscillations are damped out for all higher harmonics 
 in the limit $t \to \infty$.  

\section{Post-quench dynamics of open systems}
 \label{pqc}
 
  \begin{figure}[t]
    \centering    \includegraphics[width=\linewidth]{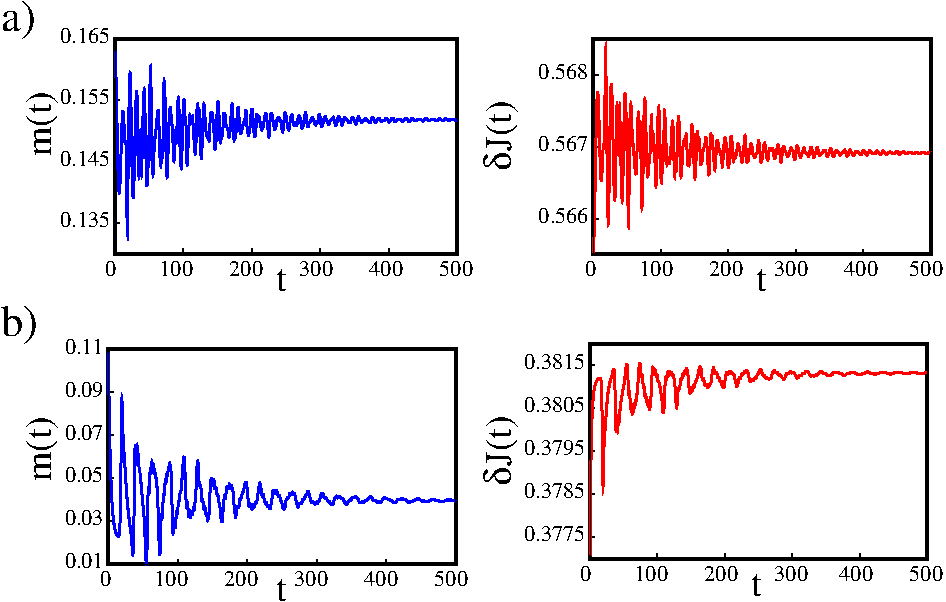}
    \caption{
    Post-quench dynamics of $m(t)$ (left) and $\delta J(t)$ (right column)
 vs time $t$ for an open GN chain with $J=1, \gamma=0.01, L=100, k_BT=0.05$, for
    a $t=0$ quench from $g_i=1$ to (a) $g_f=0.9$ and (b) $g_f=0.6$.  These results were obtained from Eqs.~\eqref{LME.3} and \eqref{LME.4}.
    } 
    \label{fig8}
\end{figure}

  \begin{figure}[t]
    \centering    
    \includegraphics[width=\linewidth]{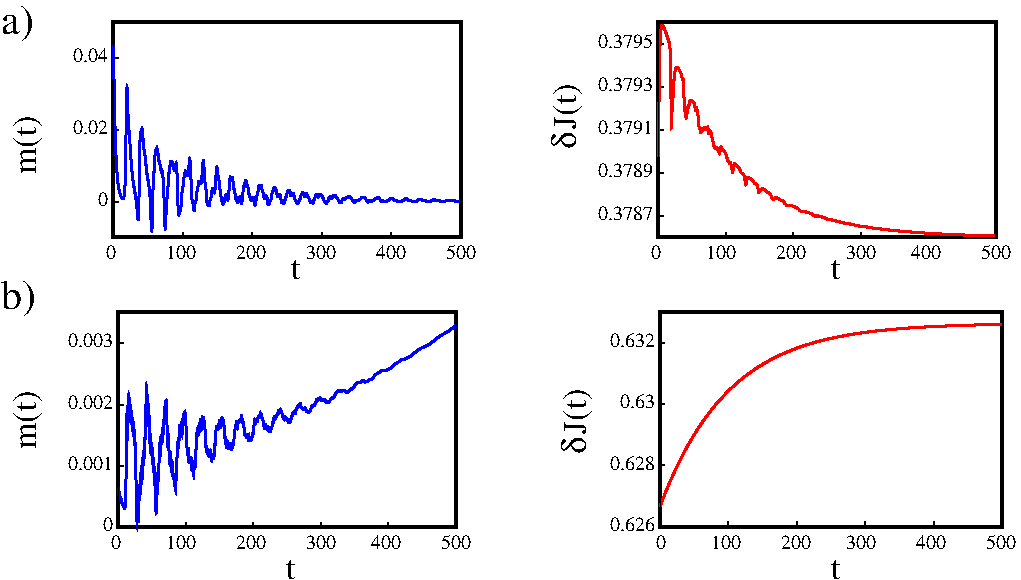}
    \caption{ (a) Post-quench dynamics of $m(t)$ (left) and $\delta J(t)$ (right)
 vs time $t$ for an open GN chain with $J=1, \gamma=0.01, L=100, k_B T=0.2$, for
    a quench from $g_i=1$ to $g_f=0.6$, see Fig.~\ref{fig8}(b) for the corresponding case with $k_B T=0.05$.  (b)  Same as (a) but for a quench from $g_i=0.6$ to $g_f=1$.
    These results were obtained from Eqs.~\eqref{LME.3} and \eqref{LME.4}. }
    \label{fig9}
\end{figure}
  
 To illustrate the relaxation dynamics of the open system with finite $\gamma>0$, we show the self-consistent order parameter dynamics as computed from Eqs.~\eqref{LME.3} and \eqref{LME.4}  in Figs.~\ref{fig8} and \ref{fig9} for fixed system size $L=100$.  In Fig.~\ref{fig8}, we address a rather low temperature, $k_B T=0.05J$, while in Fig.~\ref{fig9}, we consider the elevated temperature $k_B T=0.2J$  for a quench from the ordered to the disordered phase, see Fig.~\ref{fig9}(a), as well as from the 
 disordered to the ordered phase in Fig.~\ref{fig9}(b). Note that in the latter case, we have introduced a tiny initial mass at $t=0^+$ in order to drive 
 the post-quench ${\bf Z}_2$ symmetry breaking toward the ordered phase.
   We recall that the Lindblad approach for $\gamma>0$ is valid at not too low temperatures and for weak system-environment coupling due to the Born-Markov approximation needed for deriving the LME. 
  Specifically, in order to meet the standard requirements for the validity of the approximation \cite{Breuer2007,Ackermann2023,Zatsarynna2024},  we put $\gamma=0.01J$, that is
 $1/100$ of the typical scale of the high-frequency cutoff and at least one order of magnitude smaller than $2 \pi k_B T$. 
 Apparently, in all cases, oscillations are now damped out at long times, and $m(t)$ and $\delta J(t)$ approach their equilibrium values
 for $g=g_f$ as $t\to \infty$. A similar relaxation dynamics 
 but toward the disordered phase ($m_f=0$) is found for the elevated temperature $k_BT=0.2J$ in Fig.~\ref{fig9}(a), see the equilibrium phase diagram in Fig.~\ref{fig1}.  Again we observe damped oscillations in $m(t)$ and $\delta J(t)$ approaching their respective asymptotic values (which now vanish) at $t\to \infty$.   
 Finally, in Fig.~\ref{fig9}(b), we show $m(t)$ and $\delta J(t)$ for the case where the quench takes the system from the disordered to 
 the ordered phase at given $T$. Remarkably, the post-quench time evolution toward a finite asymptotic value of $m(t)$ suggests the interesting possibility of 
 employing a quantum quench to trigger the onset, in real time, of the ordered phase from the disordered phase.

\begin{figure}[t]
\centering    \includegraphics[width=0.8\linewidth]{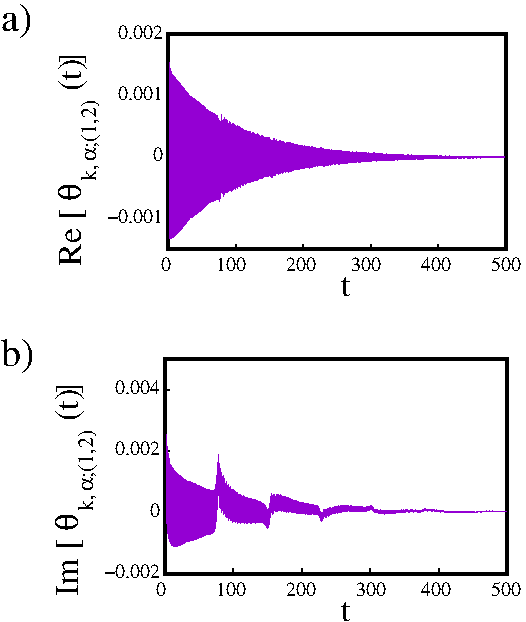}
\caption{Post-quench dynamics of the (a) real part and (b) imaginary part 
of $\theta_{k_*,\alpha;(1,2)} (t)$ vs time $t$ in the open GN system with $J=1, \gamma = 0.01, L=400, k_BT=0.05,$ after a quench from $g_i=1$ to $g_f=0.5$, for the momentum $k_* = 15 \frac{2 \pi}{L}$. These results were obtained from Eqs.~\eqref{LME.3} and \eqref{LME.4}. } 
    \label{fig10}
\end{figure}
 
In contrast to the closed system in Sec.~\ref{isopost}, 
a finite system-environment coupling $\gamma>0$ is expected to ensure a true relaxation of the open system  toward a stationary equilibrium state where all correlation matrix elements become stationary for $t\to \infty$. To verify this expectation, in Fig.~\ref{fig10}, we show the dynamics of $\theta_{k_*,\alpha;(1,2)}  (t)$ for the parameters used in Fig.~\ref{fig7} but now with $\gamma = 0.01J$. The observation of damped oscillations for $t\to \infty$ confirms the existence of a stationary equilibrium state determined by the 
post-quench system parameters for $\gamma > 0$.

The dissipative dynamics of time-dependent observables may again be described by the simplified non-self-consistent approach, see Eq.~(\ref{anan.2}) in App.~\ref{ananon}.  From our explicit solution for $\vec{\cal T}_{k,\alpha} (t)$ in Eq.~(\ref{anan.4}), with the asymptotic solution $\vec{\cal T}_{k,\alpha}$ for $t \to \infty$ in Eq.~(\ref{anan.6}), we indeed evidence 
how a finite $\gamma$ triggers the relaxation of  $\vec{T}_{k,\alpha}(t)$ toward the stationary equilibrium state.  In contrast to the self-consistent theory, however, the non-self-consistent solution predicts a damping of time-dependent observables on the time scale $\tau \sim (2 \gamma)^{-1}$. The self-consistent solution, see, e.g., Fig.~\ref{fig8}, instead exhibits a much slower attenuation of $m (t)$ and $\delta J(t)$,  due to the fact that, as evidenced in Eq.(\ref{LME.3}), 
the self consistency induces an explicit dependence on time in  the jump operators. This makes
the relaxation rate effectively depend on time and, eventually,  close to the final state 
(that corresponds to an extremal point of the 
post-quench (free) energy) the curve describing the relaxation pattern in time of 
the system tends to flatten, corresponding to an effective slowing down of the relaxation
rate. Thus, we conclude that the non-self-consistent approach here fails to yield the proper relaxation time scales, 
even though the qualitative dynamical behavior is correctly captured.

An important observation concerns the fact that, on inserting into Eq.~(\ref{mh.1}) the mean-field solution (\ref{deltadecomp}) for the order parameter, one gets back the 1D Su-Schrieffer-Heeger (SSH)  Hamiltonian \cite{Su1979}. 
At equilibrium, at finite $m$, the SSH model lies within a nontrivial topological insulator phase. A natural question is if, and possibly how, the nontrivial topology of the model appears in the post-quench relaxation dynamics 
of the system. In Ref.~\cite{Nava2024}, three of us have studied the interplay between dynamical post-quench evolution and nontrivial 
topological properties for a 2D topological superconductor taken across a dynamical phase
transition between a topologically trivial and a topological  phase. To monitor the system across the dynamical phase transition, we proposed to study response functions such as the spin-Hall conductance, which in equilibrium,  within each phase, is proportional to a topological invariant (the Chern number $C$, with $C=0$ in the trivial phase and 
$C=\pm 1$ in the topological phase). However, in contrast to the Chern number, the spin-Hall conductance remains well defined across the dynamical phase transition, which eventually allowed us to continuously monitor the system across the transition itself. The same strategy could be applied here by substituting the spin-Hall conductance with a response function suitable for a 1D system, e.g., the charge polarization \cite{Qi2008}.
In principle, one might repeat the same derivation as in Ref.~\cite{Nava2024} for a 2D system  and eventually resort to our 1D system by means
of a dimensional reduction, see Ref.~\cite{Qi2008}.  
Since here we discuss examples of relaxation dynamics without crossing a dynamical phase transition, we may infer a time evolution of the charge polarization (or some alternative response function), which on average is a monotonically decreasing (increasing) function of time when going from the ordered (disordered) to the disordered (ordered) phase, possibly with superimposed transient oscillations. Eventually, along the relaxation dynamics, the charge polarization is expected to continuously interpolate between the topological (trivial) and the trivial (topological) phase.

\section{Concluding remarks}
\label{concl}
 
 By means of a systematic application of the LME complemented with 
 the time-dependent SCMF approximation for the order parameter, we have 
studied the post-quench dynamics of a lattice version of the 1D GN model. 
In particular, we have compared the dynamics of the closed many-body quantum system 
to the open case, where the system is weakly coupled to an environment.  We 
have highlighted the importance of synoptically considering 
the dynamics of all finite-momentum correlation matrix elements. While for large quench amplitudes and in the thermodynamic limit $L \to \infty$, the order parameter dynamics of the closed system is indistinguishable from a simple relaxation toward a final equilibrium state, 
the finite-momentum correlation matrix elements still exhibit fast undamped oscillations characteristic of the underlying persistent nonequilibrium dynamics. Only when including the system-environment coupling $\gamma>0,$ the system relaxes to an equilibrium state where all possible observables become stationary.
Our results thereby shed light on the mechanisms determining the post-quench dynamics 
of quantum many-body systems. In general, when monitoring only global observables, e.g., uniform order parameters, these observables may exhibit relaxation behavior even for a closed system while other observables do not.  In open systems, however, a finite system-environment coupling $\gamma>0$ ensures the existence of an equilibrium state where all observables become stationary in the limit $t\to \infty$.  
We note that our SCMF-LME approach 
 makes no assumptions about the final state.  We find that, for $t\to \infty$, the system is  always driven  to the proper  equilibrium state determined by the post-quench
 parameters. This fact also enables the construction of the equilibrium phase diagram, 
 see Fig.~\ref{fig1}, which is consistent to previous results obtained through large-$N$ effective thermodynamic potentials  \cite{Wolff1985}. 
 By engineering the system-environment coupling and the quench protocol, one can then prepare
 a wide class of target states.  Even though we have examined a specific model, given that our conclusions are consistent with the results of previous work on different systems  \cite{Peronaci2015,Nava2023_s,Nava2024}, we are confident that our approach and conclusions apply to many other quantum many-body systems.

 \section*{Data availability}

The data underlying the figures in this work can be found at Zenodo \cite{Zenodo}.

 \begin{acknowledgments} 
We acknowledge funding by the Deutsche Forschungsgemeinschaft (DFG, German Research Foundation) under Projektnummer 277101999 - TRR 183 (project  B02), under Projektnummer EG 96/14-1, and under Germany's Excellence Strategy - Cluster of Excellence Matter and Light for Quantum Computing (ML4Q) EXC 2004/2 - 390534769.
\end{acknowledgments} 

\appendix
 \section{Equilibrium free energy}
\label{freen}

We here derive the equilibrium free energy of the GN model, see Eq.~\eqref{mh.1}, within the SCMF approximation in 
Eq.~(\ref{mh.5}), assuming spatially uniform order parameters $m$ and $\delta J$, see Eq.~\eqref{deltadecomp}.  
Written as imaginary-time functional integral \cite{Weiss2021}, with $j=1,\ldots,L$ and $\alpha=1,\ldots,N$, the partition function takes the form  
\begin{widetext}
 \begin{equation} \label{parf.3}
{\cal Z} = \int    {\cal D} \left[ \Delta_j, \bar c_{j,\alpha} ,c^{}_{j,\alpha}\right]  
\exp   \int_0^\beta  d \tau \left( -\frac{N}{2g} \sum_{j} \Delta^2_j (\tau) + \sum_{j,\alpha} \left[
\bar c_{j,\alpha}  (\tau) \partial_\tau c_{j,\alpha} (\tau)+ [J+\Delta_j(\tau)] \left(\bar c_{j,\alpha}     c_{j+1,\alpha}  + \bar c_{j+1,\alpha}  c_{j,\alpha} \right)_\tau \right] \right)
\end{equation}
\end{widetext}
with fermionic Grassmann fields $\{\bar c_{j,\alpha}(\tau),c_{j,\alpha}(\tau)\}$ and the displacement field $\Delta_j (\tau)$.  Using the SCMF approximation,  we next substitute $\Delta_j (\tau)$ by Eq.~\eqref{deltadecomp}.  While 
the uniform contribution $\delta J$ renormalizes $J\to {\cal J} = J + \delta J$,  
the staggered component $m$ is   accounted for by switching from $c_{j,\alpha}(\tau)$ (and likewise for $\bar c_{j,\alpha}$) 
to the spinor fields $c_{k,i\omega_n,\alpha,a}$ with $a=1,2$, see Eq.~\eqref{mh.6}, where the quasi-momentum $k$ covers 
only half of the Brillouin zone, $0\le k\le \pi$ with discrete momenta $k$ for finite $L$. With integer $n$, 
the fermionic Matsubara frequencies are $\omega_n = \frac{2 \pi}{\beta} \left( n+ \frac{1}{2} \right)$. 
In frequency-momentum representation, we then obtain
\begin{equation}
  {\cal Z} =  \int {\cal D} [\bar c , c] \,
e^{  -  N L \beta [m^2+ (\delta J)^2]/2 g^2 + {\cal S}[ \bar c , c]} .
\label{parf.5}
\end{equation}
Defining the bispinor $C_{k,i\omega,\alpha}= (c_{k,i\omega,\alpha,1}  , c_{k,i\omega,\alpha,2})$, we obtain the fermionic action   
\begin{equation} \label{parf.6}
  {\cal S} [\bar c,c] = \frac{1}{\beta L}\sum_{\alpha,\, 0\le k \le\pi, \,i\omega_n}     \bar C_{k,i\omega_n,\alpha}
{\cal M}_{k,i\omega_n} C_{k,i\omega_n,\alpha} 
\end{equation}
with 
\beq
{\cal M}_{k,i\omega} = \left( \begin{array}{cc}  i \omega + 2 {\cal J} \cos k   & -2im \sin k \\
2im \sin k &  i \omega - 2 {\cal J}   \cos k   \end{array}\right). 
\label{parf.7}
\eneq
Integrating over the fermion fields, the free energy density  
$f = -(k_B T/L) \ln {\cal Z}$ follows as
\beq
f =\frac{[m^2 + (\delta J)^2] N }{2 g } - \frac{N k_BT}{L}\sum_{k,i\omega_n} 
{\rm Tr}\ln {\cal M}_{k,i\omega_n}  . 
\label{parf.8}
\eneq
Diagonalizing ${\cal M}_{k,i\omega}$ in Eq.~\eqref{parf.7}, we obtain 
\begin{eqnarray}
f &=&  \frac{[m^2 + (\delta J)^2]N }{2 g } \label{exp.1}  \\ \nonumber &-& \frac{Nk_B T}{  L} \sum_{k, i\omega_n} 
\ln [ - \omega_n^2- 4 ( {\cal J}^2 \cos^2 k + m^2 \sin^2 k )].
\end{eqnarray} 
The self-consistency equations for $m$ and $\delta J$ are determined by 
$\frac{\partial f}{\partial \delta J} = \frac{\partial f}{\partial m} = 0$. 
As a result, we arrive at Eq.~(\ref{mh.10}). Taking the thermodynamic limit $L\to \infty$, the self-consistency relation for $m$ is given by
\begin{equation}\label{ap.a.x1}
m =\frac{4gm}{2\pi} \int_0^\pi d k \,\frac{\sin^2 k}{\epsilon_k} \tanh \frac{\epsilon_k}{2k_B  T}, 
\end{equation}
with $\epsilon_k$ in Eq.~\eqref{dispferm}.
We note that by allowing for a finite chemical potential $\mu$  and
computing  $\frac{\partial f(\mu=0)}{\partial \mu}$,  the average electronic occupation $\bar{n}$ follows as expected for the half-filled case,
\beq
\bar{n}=    \frac{1}{\beta L} \sum_{k, i\omega_n} 
 \frac{2  i\omega_n  }{\omega_n^2 + \epsilon_k^2} = \frac{1}{2}   . 
\label{xx.e2}
\eneq
The phase boundary in the $g$-$T$ plane, see Fig.~\ref{fig1}, follows from the condition $m(g,T)=0$, where the critical curve $T=T_*(g)$ 
separates the ordered ($m\ne 0$) from the disordered ($m=0$) phase.
 For $T \to 0$, nontrivial solutions $m=\pm m_*$ with $m_*\ll {\cal J}$ follow from Eq.~\eqref{ap.a.x1} 
by expanding $\epsilon_k$ near the band minimum at $k=\frac{\pi}{2}$. Writing $k = \frac{\pi}{2} + q$ with $|q|\ll \Lambda\sim \frac{\pi}{2}$, we find 
$\epsilon_k\approx 2 {\cal J} \sqrt{ q^2 + \left( \frac{m}{{\cal J}} \right)^2 }$.
Using $\bar n=1/2$, we obtain   
\beq
 m_*  \simeq \pi {\cal J}  e^{-\frac{\pi {\cal J}}{2g}}. 
\label{yy.x7}
\eneq
With increasing temperature, the system (in the large-$N$ limit) undergoes a second-order phase transition 
toward the disordered ($m=0$) phase at a critical temperature $T_*\simeq m_*/k_B$  \cite{Wolff1985}.  Note that $T_*$ increases
when increasing the interaction strength $g$ at fixed ${\cal J}$. 

\section{LME and dynamics of correlations}
\label{LME_appe}

We here provide details on the LME for the density matrix $\rho(t)$.  The LME also determines the dynamics of the correlation matrix in our quasi-free fermion system. 
In particular, we sketch the derivation of the LME from a microscopic model for the system-bath interaction by assuming a fermionic reservoir, e.g., a metallic gate tunnel-coupled to the 1D GN chain \cite{Nava2024,Nava2025_s}. 
Remarkably, the resulting order parameter dynamics is
equivalent to previous semi-phenomenological results for the dissipative
order parameter dynamics in a BCS superconductor \cite{Yuzbashyan2005,Cui2019}. 
Since these equations involve effects beyond BCS theory, e.g., quasi-particle interactions and/or interactions between quasiparticles and order parameter fluctuations, our time-dependent SCMF approach can capture effects beyond simple time-independent mean-field theories \cite{Yuzbashyan2005,Peronaci2015,Cui2019,Nava2024}. 

We start from the total Hamiltonian $H_{\rm tot}=H_S+H_B+H_T$, where $H_S$
describes the many-body fermion system of interest, $H_B$ corresponds to a thermal fermionic environment, and $H_T$ models the weak system-environment coupling. 
The system and bath fermion annihilation operators in momentum space (with flavor index $\alpha$) are denoted by $c_{k,\alpha}$ and $d_{k,\alpha}$, respectively. We use the standard tunneling Hamiltonian \cite{Weiss2021} with flavor-independent 
tunnel amplitudes $t_k$,
$H_T= \sum_{k,\alpha}  t_k  c_{k,\alpha}^\dagger d^{}_{k,\alpha}+ {\rm h.c.},$
assuming an extended tunnel contact such that 
translation invariance along the chain direction is preserved.  
The fermionic reservoir is modeled as free Fermi gas with dispersion $\xi_k$, 
$H_B  = \sum_{k,\alpha} \xi_k d_{k,\alpha}^\dagger d_{k,\alpha}^{}.$
Following the standard LME derivation \cite{Breuer2007}, we assume that $H_T$ is turned on at 
$t=0$ and consider the time evolution of the density matrix of the total system, $\rho_{\rm tot}(t)$, in the interaction representation. To leading order in $H_T$, i.e., using the Born approximation, one obtains \cite{Breuer2007}
\beq
\frac{d \rho_{\rm tot} (t)}{dt} \approx - i \int_0^t d t'  [H_T (t) , [H_T (t') , \rho_{\rm tot} (t) ]] . 
\label{appe.c.4}
\eneq
We then employ the Markov approximation, assuming that the bath memory time is 
very short. With the equilibrium bath density matrix  $\rho_B$, we then 
have $\rho_{\rm tot} (t) = \rho (t) \otimes \rho_B$.  
Integrating Eq.~(\ref{appe.c.4}) over the $d$ fermions and using the Schr\"odinger picture for $\rho (t)$, the LME follows as
\begin{eqnarray}
&& \frac{d \rho (t)}{d t } = - i [ H_S , \rho (t) ] + \label{appe.c.5} \\
&&+\, \gamma \sum_{k,\alpha,\lambda = \pm }  \left[ f (-\lambda \epsilon_k ) {\cal D}[\Gamma_{k,\alpha,\lambda}] \rho (t)
+ f (\lambda \epsilon_k ) {\cal D}[\Gamma_{k,\alpha,\lambda}^\dagger] \rho \right] . 
\nonumber 
\end{eqnarray}
We here neglected the $k$-dependence of $t_k$ and used the golden rule expression $\gamma=4\pi |t_k|^2$ for the hybridization scale between the system and the environment. 
In our case, $H_S=H$ is given by Eq.~(\ref{mh.7}) and the
operators $\Gamma_{k,\alpha,\lambda}$  by Eq.~(\ref{mh.8}). For $\gamma=0^+$ and time-independent Hamiltonian, the order parameters $m$ and $\delta J$ within our SCMF approach satisfy the self-consistency conditions in Eq.~(\ref{mh.10}). 
Due to the parameter quench, for $t>0$,
$m(t)$ and $\delta J(t)$ depend on time, and hence also $H (t)$ becomes time-dependent, with  instantaneous eigenenergies $\epsilon_k (t)$ and the corresponding eigenmode operators $\Gamma_{k,\alpha,\lambda} (t)$.  We thereby  obtain the LME as quoted in the main text.

For the system of quasi-free fermions considered here, the LME can be equivalently formulated for the correlation matrix $\theta_{k,\alpha;(a,a')} (t)$ in Eq.~\eqref{thetadef} \cite{Fazio2025}.
Specifically, if $\rho (t)$ satisfies Eq.~(\ref{LME.1}), we obtain the dynamical equations
in Eq.~\eqref{LME.3} for the correlation matrix elements, with the self-consistency conditions \eqref{LME.4}.  Inserting $m(t)$ and $\delta J(t)$ into the system Hamiltonian, 
we obtain $H(t)$ in Eq.~\eqref{LME.5}, which equivalently can be written as 
\beq
H(t) = \sum_{0 \leq k \leq \pi,\, \alpha} (c_{k,\alpha,1}^\dagger  , c_{k,\alpha,2}^\dagger )  \vec{{\cal H}}_k (t) \cdot \vec{\sigma}   \left(\begin{array}{c}c_{k,1,\alpha}\\ c_{k,2,\alpha}\end{array}\right) 
\label{appe.c.9}
\eneq
with the vector $\vec\sigma$ of Pauli matrices and $\vec{{\cal H}_k}(t)$ given by Eq.~\eqref{pq0.2} but with $m_f\to m(t)$ and $J_f\to {\cal J}(t)$.

From Eq.~(\ref{appe.c.9}), an alternative motivation for employing the LME  for 
the post-quench time evolution opens up. Indeed, with the isospin vector 
$\vec{\cal T}_{k,\alpha}$ in Eq.~\eqref{pq0.1}, which is subject to the initial condition \eqref{pq0.4}, we find from Eq.~(\ref{LME.1})  that the dynamics of $\vec{{\cal T}}_{k,\alpha}$ obeys a Bloch-like equation, see also Ref.~\cite{Cui2019},
\beq
\frac{d \vec{\cal T}_{k,\alpha} (t) }{ d t} = 2 \vec{\cal H}_k (t) \times \vec{\cal T}_{k,\alpha} (t) -  \gamma \vec{\cal T}_{k,\alpha} (t) 
+ \vec{\Lambda}_{k}(t), 
\label{appe.c.13}
\eneq
with the $k$- and time-dependent vector
\beq
\vec{\Lambda}_k (t) = -\gamma \left[1-2f\left(\epsilon_k(t)\right)\right] 
\left(\begin{array}{c} 0 \\  \sin \vartheta_k (t)  \\ \cos  \vartheta_k (t)  \end{array}\right).
\label{appe.c.14}
\eneq
The angles $\vartheta_k(t)$ are defined in Eq.~\eqref{mh.9} with $m\to m(t)$ and ${\cal J}\to {\cal J}(t)$. 
For $\gamma = 0$, Eq.~(\ref{appe.c.13}) reduces to a Bloch equation for the ``spin'' vector 
$\vec {\cal T}_{k,\alpha}$ in the ``external field'' $\vec {\cal H}_k (t)$. 
In fact, in the absence of a time-dependent SCMF relation linking  $m(t)$ and $\delta J(t)$, 
and therefore $\vec {\cal H}_k (t)$, to  $\vec {\cal T}_{k,\alpha} (t)$, 
the system completely decouples into a set of independent Bloch equations for each $k$ and $\alpha$.  
We note that dissipation effects $\propto \gamma$ modify the Bloch equations in Eq.~\eqref{appe.c.14} by introducing effects due to
longitudinal (inverse relaxation time $T_1^{-1}$) and transverse ($T_2^{-1}$) damping rates, with  
$T_1^{-1} = T_2^{-1} = \gamma$ from Eq.~(\ref{appe.c.13}). In addition, we explicitly  obtain 
an expression for $\vec{\Lambda}_k(t)$. For $t\to \infty,$ this vector is proportional to the equilibrium pseudospin configuration and is usually introduced {\it ad hoc}. We emphasize that the above Bloch equations coincide with similar results of previous works on related problems \cite{Yuzbashyan2006b,Cui2019}. 
 
 \section{Non-self-consistent solution}
 \label{ananon}
 
 We here discuss the analytical solution of Eq.~(\ref{appe.c.13}) in the absence of 
  self-consistency.  Specifically, we study the time dependence of 
 $m (t)$ and  $\delta J (t)$ after a quench in the interaction strength, $g_i\to g_f$, at time $t=0$. Without self-consistency, we assume 
  \begin{equation}
  m (t) =  \Theta(-t) m_i +  \Theta (t) m_f,\quad
  \delta J (t) = \Theta (-t) \delta J_i+ \Theta (t)\delta J_f  , 
  \label{anan.1}
  \end{equation}
with the Heaviside step function $\Theta$, where
 $m_{i/f}$ and $\delta J_{i/f}$ are determined from Eq.~(\ref{mh.10}) with $g=g_{i/f}$.  This approximation is expected to be accurate for small quench amplitude $|g_f-g_i|$. However, in general, it provides a useful guide to the size dependence of the frequency scales governing the post-quench dynamics.  Without time-dependent self-consistency, the post-quench dynamics of 
${\vec{\cal T}}_{k,\alpha}(t)$ is fully determined by $H$ with parameters $g_f,m_f$ and $\delta J_f$,
where the Bloch equations \eqref{appe.c.13} decouple
in both $k$ and $\alpha$. Dropping the flavor index henceforth, we obtain for $t>0$ from  
 Eq.~(\ref{appe.c.13}) the decoupled Bloch equations
\begin{equation} \frac{d}{dt} {\vec{\cal T}}_k(t) =  
\label{anan.2} {\cal B}_k\cdot {\vec{\cal T}}_k(t) +{\vec\Lambda}_k ,
\end{equation} 
with the matrix 
\begin{equation}
  {\cal B}_k = \left(\begin{array}{ccc} -  \gamma & -2 \epsilon_{k,f}\cos 
\vartheta_{k,f}  &  2 \epsilon_{k,f} \sin \vartheta_{k,f}  \\
2 \epsilon_{k,f}\cos \vartheta_{k,f}  & -  \gamma & 0 \\
- 2 \epsilon_{k,f} \sin \vartheta_{k,f}  & 0 & - \gamma \end{array}\right).
\end{equation}
Given  our approximations, both the matrix ${\cal B}_k$ and the vector $\vec{\Lambda}_k$ in Eq.~\eqref{anan.2} are time-independent. In particular, we get 
${\vec\Lambda}_k= -  \gamma  ( 0 , \sin \vartheta_{k,f} , \cos \vartheta_{k,f} )^T.$ 
With the initial condition \eqref{pq0.4}, integration of Eq.~(\ref{anan.2}) 
yields 
\begin{widetext}
\begin{equation}
 \left(\begin{array}{ccc} 0 & \sin \vartheta_{k,f}  & \cos \vartheta_{k,f}  \\
1 &   i \cos\vartheta_{k,f} & - i \sin\vartheta_{k,f}  \\ 1 &  - i \cos \vartheta_{k,f} &   i \sin \vartheta_{k,f} \end{array}\right)
\cdot {\vec{\cal T}}_k(t) 
= \left(\begin{array}{c} - e^{- \gamma t} \cos (\vartheta_{k,f} - \vartheta_{k,i} ) + 
\frac{1-e^{- \gamma t}}{ \gamma}  [ \sin (\vartheta_{k,f} ) \Lambda_k^y  + 
 \cos (\vartheta_{k,f} ) \Lambda_{k}^z ] \\ i e^{- ( \gamma - 2i \epsilon_{k,f} ) t} \sin (\vartheta_{k,f} - \vartheta_{k,i} )  +
 i  \frac{1-e^{ -(\gamma - 2i \epsilon_{k,f} )t}}{\gamma - 2i \epsilon_{k,f}} 
  [   \cos (\vartheta_{k,f} )  \Lambda_{k}^y  - \sin (\vartheta_{k,f} ) \Lambda_k^z ] \\  
 - i e^{-( \gamma +  2i \epsilon_{k,f} ) t} \sin (\vartheta_{k,f} - \vartheta_{k,i} )  
 -i  \frac{1-e^{- (\gamma+ 2i \epsilon_{k,f} )t}}{\gamma+ 2i \epsilon_{k,f}} 
[  \cos (\vartheta_{k,f} )  \Lambda_k^y  - \sin (\vartheta_{k,f} ) \Lambda_k^z ]  \end{array}\right) . 
  \label{anan.4}
  \end{equation}
  \end{widetext}
Simple limiting cases correspond to (i) the case $\gamma = 0$, and to (ii) the case $t\to \infty$ for $\gamma>0$. For case (i), we obtain 
\begin{eqnarray} \label{anan.5}
T_{k}^x(t) &=& - \frac{ 2 (m_i {\cal J}_f - {\cal J}_i m_f) \sin (2k)}{\epsilon_{k,i} \epsilon_{k,f}} \sin (2 \epsilon_{k,f} t),  \\
T_{k}^y(t) &=&   \frac{8 [{\cal J}_i {\cal J}_f \cos^2 k + m_i m_f \sin^2 k ]m_f \sin k}{\epsilon_{k,i}\epsilon_{k,f}^2 } 
\nonumber \\
&-& \frac{4 (m_i {\cal J}_f-m_f{\cal J}_i) {\cal J}_f\sin(2k) \cos k }{\epsilon_{k,i} \epsilon_{k,f}^2} \cos (2 \epsilon_{k,f} t), \nonumber \\
T_{k}^z(t) &=&  \frac{8 [{\cal J}_i {\cal J}_f \cos^2 k + m_i m_f \sin^2 k ]{\cal J}_f\cos k}{\epsilon_{k,i}\epsilon_{k,f}^2} 
\nonumber \\ &-& \frac{4 (m_i{\cal J}_f-m_f{\cal J}_i) m_f\sin(2k) \sin k}{\epsilon_{k,i} \epsilon_{k,f}^2} \cos (2 \epsilon_{k,f} t).\nonumber
 \end{eqnarray}
 For case (ii), we instead find 
 \beq
\lim_{t \to \infty} {\vec{\cal T}}_k(t)  =-[1-2f(\epsilon_{k,f})]
\left(\begin{array}{c}0 \\ \sin \vartheta_{k,f}  \\ \cos \vartheta_{k,f}  \end{array}\right), 
\label{anan.6}
\eneq
in accordance with the stationary equilibrium configuration for $g=g_f$.

 \section{Dynamical evolution of the system and the Generalized Gibbs Ensemble}
\label{GGE}

In order to investigate the large-time behavior of our system, in particular in the $\gamma = 0$ limit, 
we have to consider the special role played by the quantum integrability, in combination with 
ETH (see, for instance, \cite{Yuzbashyan2016}). First of all, we note that the integrability of the 1D GN model in 
the continuum limit is, by now, well established \cite{Thies2014}. Once resorting to the lattice model Hamiltonian 
in Eq.(\ref{mh.1b}) and after the decoupling by means of the lattice displacement field in Eq.(\ref{mh.1}), integrability is 
definitely preserved within the non-self consistent mean-field theory we refer to in Appendix \ref{ananon} to 
qualitatively discuss the time dynamics of the closed system. Indeed, setting, at given values of ${\cal J}$ and 
$m$,
\beq
H_{\rm MF}  = \sum_{0\leq k \leq \pi} \sum_{\alpha=1}^N \left(c_{k,\alpha,1}^\dagger , c_{k,\alpha,2}^\dagger \right)
{\cal H}_k \left(\begin{array}{c} c_{k,\alpha,1} \\ c_{k,\alpha,2}\end{array}\right) \;\; , 
\label{apn.1}
\eneq
\noindent
with 
\beq
{\cal H}_k = \left(\begin{array}{cc} -2{\cal J} \cos k & -2im\sin k \\ 
2i m \sin k & 2{\cal J} \cos k \end{array} \right) 
\;\; , 
\label{apn.2}
\eneq
\noindent
an extensive set of local operators commuting with $H_{\rm MF}$, $\{{\cal C}_{k,\alpha}\}$,  is 
readily recovered by setting 

\begin{eqnarray}
{\cal C}_{k,\alpha} &=& -i \sin (\vartheta_{k,f} ) \{ c_{k,\alpha,1}^\dagger c_{k,\alpha,2} 
- c_{k,\alpha,2}^\dagger c_{k,\alpha,1} \} \nonumber \\
&+& \cos (\vartheta_{k,f} ) \{c_{k,\alpha,1}^\dagger c_{k,\alpha,1} - 
c_{k,\alpha,2}^\dagger c_{k,\alpha,2} \} 
\:\: , 
\label{apn.3}
\end{eqnarray}
\noindent
with $\vartheta_k$ defined in Eq.(\ref{mh.9}). In the closed system, the average values of the 
${\cal C}_{k,\alpha}$ is fixed by the initial condition. In particular, once the system has been
prepared in a state  described by the Boltzmann density matrix 
$\rho_0  = e^{-\beta H_{{\rm MF},i}} /{\rm Tr} [  e^{-\beta H_{{\rm MF},i}} ]$, with 
$H_{{\rm MF},i}$ given by Eq.(\ref{apn.1}) with $\vartheta_k \to \vartheta_{k,i}$, 
one obtains 

\begin{eqnarray}
\bar{C}_{k,\alpha } &=& \frac{ {\rm Tr} [{\cal C}_{k,\alpha} e^{-\beta H_{{\rm MF},i}} ]}{{\rm Tr} 
[e^{-\beta H_{{\rm MF},i}}]} \nonumber \\
 &=& \cos [\vartheta_{k,i} - \vartheta_{k,f} ] [-1+2f(\epsilon_{k,i})]
\:\: . 
\label{apn.4}
\end{eqnarray}
\noindent
As discussed in detail in \cite{Rigol2007,Alessio2016}, Eqs.(\ref{apn.4}) determine the asymptotic ($t \to \infty$) state of 
the system in terms of a Generalized Gibbs Ensemble (GGE) density matrix, $\rho_{\rm GGE}$. 
In particular, $\rho_{\rm GGE}$ is determined by introducing a set of Lagrange multipliers, 
$\{\lambda_{k,\alpha}\}$, determined by requiring that, when computing the average values of 
${\cal C}_{k,\alpha}$ in the asymptotic state, one obtains the same results as in 
Eqs.(\ref{apn.4}). Specifically, one sets 

\beq
\rho_{\rm GGE} = \frac{e^{-\sum_{k,\alpha} \lambda_{k,\alpha} {\cal C}_{k,\alpha}}}{{\rm Tr} 
[ e^{-\sum_{k,\alpha} \lambda_{k,\alpha} {\cal C}_{k,\alpha}} ]} 
\:\: , 
\label{apn.5}
\eneq
\noindent
with the $\lambda_{k,\alpha}$ determined as specified above. Note that, to reproduce the exact values 
of the constant of motion for every $k$, the GGE must incorporate $k$-dependent 
Lagrange multipliers, reflecting the system's integrability.  Importantly, when 
writing ${\cal C}_{k,\alpha}$ in terms of the eigenmodes $\Gamma_{k,\alpha,\pm,f}$ defined 
in Eq.(\ref{mh.8}) with $\vartheta_k = \vartheta_{k,f}$,  one gets 

\beq
{\cal C}_{k,\alpha} = \Gamma_{k,\alpha,+}^\dagger \Gamma_{k,\alpha,+} - \Gamma_{k,\alpha,-}^\dagger 
\Gamma_{k,\alpha,-}
\;\; . 
\label{apn.6}
\eneq
\noindent
Given Eq.(\ref{apn.6}) and the fact that, in terms of the energy eigenmodes, one gets, 
for the post-quench Hamiltonian 

\begin{eqnarray}
H_{{\rm MF},f} &=& \sum_{0 \leq k \leq \pi} \sum_{\alpha=1}^N \epsilon_{k,f} \{\Gamma_{k,\alpha,+}^\dagger 
\Gamma_{k,\alpha,+} - \Gamma_{k,\alpha,-}^\dagger \Gamma_{k,\alpha,-} \}  \nonumber \\
&=& \sum_{0 \leq k \leq \pi} \sum_{\alpha=1}^N \epsilon_{k,f} {\cal C}_{k,\alpha} 
\;\; , 
\label{apn.7}
\end{eqnarray}
one readily concludes that in any state described by the density matrix corresponding to 
the microcanonical or to the canonical ensemble at temperature $T$ (such as $\rho_0$ with 
$H_{{\rm MF},i} \to H_{{\rm MF},f}$), one would get $\bar{C}_{k,\alpha} = -1+2f(\epsilon_{k,f})$, which would 
not be consistent with Eqs.(\ref{apn.3}) for an equilibrium thermodynamical state with an uniquely 
defined temperature and with the conservation of $\bar{C}_{k,\alpha}$. 
Indeed, $\rho_{\rm GGE}$ corresponds to a peculiar, nonequilibrium state, as witnessed
by our analysis of Section \ref{isopost}. 

When implementing the full time-dependent SCMF approach, the $H_{\rm MF}$ takes an explicit 
dependence on time, which does no longer allow for recovering simple expressions for 
the constants of motion, as we did without accounting for self-consistency. Yet, we may infer 
again that, as $t \to \infty$, the system is described by a GGE, with pertinent values of 
the $\bar{C}_{k,\alpha}$ that can be recovered by, for instance, 
approximating the whole time evolution with a sequence of steps within intervals of 
time in which $m(t)$ and ${\cal J} (t)$ are constant. While, over all, we conclude that
we can only numerically recover the values of the $\bar{C}_{k,\alpha}$,  yet, the striking 
qualitative analogy between the behavior of the correlation functions that we compute
in Section \ref{isopost} with, and without, accounting for self-consistency, suggests that 
in both cases the system evolves toward a GGE, that is, toward an out of equilibrium state,
determined by the initial values of the constants of motion ${\cal C}_{k,\alpha}$. 

The situation is completely different at a finite $\gamma$. Referring, again, first to
the non self-consistent solution for the equations of motion, from Eq.(\ref{anan.4})
we first infer that now the $\bar{C}_{k,\alpha}$ explicitly depend on $t$ and, therefore, that
now they are no longer constants of motion. Moreover, Eq.(\ref{anan.6}) shows that, 
as $t \to \infty$, one gets that $ \lim_{t \to \infty} \bar{C}_{k,\alpha} (t) = -1+2f(\epsilon_{k,f} )$, 
that is the value corresponding to the thermodynamical equilibrium state of the
system described by the Hamiltonian $H_{{\rm MF},f}$ at temperature $T$. This is a 
``true'' equilibrium state, only reached by turning on a nonzero coupling $\gamma$ to the 
thermal bath.

\bibliography{biblio_revised}
\end{document}